\documentclass[twocolumn]{aastex63}

\usepackage{color,graphicx,natbib}
\usepackage{graphicx}
\usepackage{amsmath}
\usepackage{amssymb}
\usepackage{bm}

\def\ra#1#2#3{#1$^{\rm h}$ #2$^{\rm m}$ #3$^{\rm s}$}
\def\dec#1#2#3{$#1^\circ #2' #3''$}

\shorttitle{Late-time Radio Short GRBs}
\shortauthors{Schroeder et al.}

\begin{document}

\author[0000-0001-9915-8147]{Genevieve~Schroeder}
\affiliation{Center for Interdisciplinary Exploration and Research in Astrophysics and Department of Physics and Astronomy, Northwestern University, 2145 Sheridan Road, Evanston, IL 60208-3112, USA}

\author[0000-0001-8405-2649]{Ben~Margalit}
\thanks{NASA Einstein Fellow}
\affiliation{Astronomy Department and Theoretical Astrophysics Center, University of California, Berkeley, Berkeley, CA 94720, USA}

\author[0000-0002-7374-935X]{Wen-fai~Fong}
\affiliation{Center for Interdisciplinary Exploration and Research in Astrophysics and Department of Physics and Astronomy, Northwestern University, 2145 Sheridan Road, Evanston, IL 60208-3112, USA}

\author[0000-0002-4670-7509]{Brian~D.~Metzger}
\affiliation{Center for Computational Astrophysics, Flatiron Institute, 162 W. 5th Avenue, New York, NY 10011, USA}
\affiliation{Department of Physics and Columbia Astrophysics Laboratory, Columbia University, New York, NY 10027, USA}

\author[0000-0003-3734-3587]{Peter~K.~G.~Williams}
\affiliation{Center for Astrophysics {\textbar} Harvard \& Smithsonian,
 60 Garden St., Cambridge, MA 02138, USA}
\affiliation{American Astronomical Society,
 1667 K Street NW, Suite 800, Washington, DC 20006, USA}

\author[0000-0001-8340-3486]{Kerry~Paterson}
\affiliation{Center for Interdisciplinary Exploration and Research in Astrophysics and Department of Physics and Astronomy, Northwestern University, 2145 Sheridan Road, Evanston, IL 60208-3112, USA}

\author[0000-0002-8297-2473]{Kate~D.~Alexander}
\thanks{NASA Einstein Fellow}
\affiliation{Center for Interdisciplinary Exploration and Research in Astrophysics and Department of Physics and Astronomy, Northwestern University, 2145 Sheridan Road, Evanston, IL 60208-3112, USA}

\author[0000-0003-1792-2338]{Tanmoy~Laskar}
\affiliation{Department of Physics, University of Bath, Claverton Down, Bath, BA2 7AY, United Kingdom}

\author[0000-0001-9652-8384]{Armaan~V.~Goyal}
\affiliation{Center for Interdisciplinary Exploration and Research in Astrophysics and Department of Physics and Astronomy, Northwestern University, 2145 Sheridan Road, Evanston, IL 60208-3112, USA}

\author[0000-0002-9392-9681]{Edo~Berger}
\affiliation{Center for Astrophysics {\textbar} Harvard \& Smithsonian,
 60 Garden St., Cambridge, MA 02138, USA}

\title{A Late-time Radio Survey of Short GRBs at $z<0.5$: New Constraints on the Remnants of Neutron Star Mergers}

\begin{abstract}
    Massive, rapidly-spinning magnetar remnants produced as a result of binary neutron star (BNS) mergers may deposit a fraction of their energy into the surrounding kilonova ejecta, powering a synchrotron radio signal from the interaction of the ejecta with the circumburst medium. Here, we present 6.0~GHz Very Large Array (VLA) observations of nine, low-redshift short gamma-ray bursts (GRBs; $z<0.5$) on rest-frame timescales of $\approx 2.4-13.9$~yr following the bursts. We place $3\sigma$ limits on radio continuum emission of $F_{\nu}\lesssim 6-20\,\mu$Jy at the burst positions, or $L_{\nu} \lesssim (0.6-8.3) \times10^{28}$~erg~s$^{-1}$~Hz$^{-1}$. Comparing these limits with new light curve modeling which properly incorporates relativistic effects, we obtain limits on the energy deposited into the ejecta of $E_{\rm ej}\lesssim (0.6-6.7)\times 10^{52}$~erg ($E_{\rm ej}\lesssim (1.8-17.6) \times 10^{52}$~erg) for an ejecta mass of $0.03\,M_{\odot}$ ($0.1\,M_{\odot}$). We present a uniform re-analysis of 27 short GRBs with $5.5-6.0$~GHz observations, and find that $\gtrsim 50\%$ of short GRBs did not form stable magnetar remnants in their mergers. 
    Assuming short GRBs are produced by BNS mergers drawn from the Galactic BNS population plus an additional component of high-mass GW194025-like mergers in a fraction $f_{\rm GW190425}$ of cases, we place constraints on the maximum mass of a non-rotating neutron star (NS) ($M_{\rm TOV}$), finding $M_{\rm TOV} \lesssim 2.23\,M_{\odot}$ for $f_{\rm GW190425}=0.4$; this limit increases for larger values of $f_{\rm GW190425}$.
    The detection (or lack thereof) of radio remnants in untargeted  surveys such as the VLA Sky Survey (VLASS) could provide more stringent constraints on the fraction of mergers that produce stable remnants. If $\gtrsim 30-300$ radio remnants are discovered in VLASS, this suggests that short GRBs are a biased population of BNS mergers in terms of the stability of the remnants they produce.
\end{abstract}


\section{Introduction}

The detection of gravitational waves (GWs) has enabled the first definitive discoveries of binary neutron star (BNS) mergers \citep{gw170817,gw190425}.  Of particular interest is the nature of the neutron star (NS) remnant produced by such mergers (e.g., \citealt{Bernuzzi20}) and how long it survives after the coalescence before collapsing to a black hole (BH), as this is directly tied to the luminosity and evolution of the resulting electromagnetic signature (e.g.~\citealt{MargalitMetzger2019}).  Among the many open questions is whether BH formation is always requisite for the production for an ultra-relativistic short gamma-ray burst (GRB) jet (e.g.~\citealt{MurguiaBerthier+17}), or whether a long-lived magnetized NS remnant (``magnetar'') could also be the engine of some bursts (e.g.~\citealt{Zhang&Meszaros01,Metzger+08,Moesta+20}). 

The nature of the merger remnant is particularly sensitive to the total initial mass of the binary, and indeed the first two BNS mergers, GW170817 and GW190425, had distinct total masses of $2.74^{+0.04}_{-0.01}\,M_{\odot}$ and $ 3.4^{+0.3}_{-0.1}\,M_{\odot}$ \citep{gw170817,gw190425}, supporting the existence of diverse properties of the merger remnants and resulting EM signatures. The existence of an indefinitely stable, {\it hyper}- or {\it supra}-massive NS remnant (where the difference depends on whether differential rotation is required to stabilize the remnant; e.g.~\citealt{Shibata&Taniguchi06}), depends sensitively on the total mass of the system relative to various threshold masses which scale with the Tolman–Oppenheimer–Volkoff mass, $M_{\rm TOV}$ (the maximum stable mass of a cold non-rotating NS). The most massive binaries are expected to undergo prompt collapse to a BH (e.g.~\citealt{Bauswein+13}), while lower mass binaries can remain stable for timescales significantly longer, until the critical amount of angular momentum is removed via magnetic dipole spin-down.  In some cases where the mass of the binary is sufficiently low, the merged object may remain indefinitely stable as a NS, even once it has spun down completely (e.g.~\citealt{Giacomazzo&Perna13}).  

The value of $M_{\rm TOV}$ is uncertain observationally but is of particular interest because it is sensitive to the unknown equation of state of the NS \citep{Lattimer&Prakash16}.  Lower limits on $M_{\rm TOV}$ are available from the masses of pulsars, with the most constraining at present being $M_{\rm TOV}>2.14^{+0.09}_{-0.10}M_{\odot}$ from the mass of J0740+6620 \citep{Cromartie+20}. However, BNS mergers offer one of the few potential probes of {\it upper} limits on $M_{\rm TOV}$ (e.g.~\citealt{Lawrence+15,MargalitMetzger17,Shibata+19}).

If the merger remnant survives even for a brief period of time, the additional reservoir of rotational energy $-$ if coupled effectively to its surroundings $-$ will substantially boost the kinetic energy of the merger ejecta.  As the latter interacts with the circum-merger environment, it will decelerate via a shock, generating synchrotron emission that is predicted to peak at $\sim$GHz frequencies on $\sim$year timescales, depending on the properties of the shock and the environment \citep{NakarPiran2011}.  As pointed out by \citet{MetzgerBower2014}, this signal could be greatly enhanced in the case of a long-lived magnetar remnant relative to mergers that result in prompt BH formation.  The existence of a long-lived NS remnant can also have a significant effect on the color and evolution of the resulting ``kilonova'' signature \citep{yu+13,Metzger&Fernandez14,Kasen2015}, as well as the early-time X-ray signatures of the merger (e.g.~\citealt{Metzger&Piro14}), which can in principle be probed by follow-up observations of GW events. 

Cosmological short-duration gamma-ray bursts (GRBs), which are detected to $z \approx 2$ \citep{Fong2017} and originate from BNS (and/or possibly NS-BH) mergers \citep{Berger2014,gw170817grb,Gompertz2020}, provide a promising route to place constraints on the nature of the remnant. They have been discovered and well-localized since 2004 by NASA's Neil Gehrels {\it Swift} Observatory \citep{Gehrels2004} and provide the necessary long time baseline required to match the $\approx 1-10$~year peak timescales of the predicted radio signatures. Indeed, previous radio studies of short GRBs on year to several year timescales after the events have endeavored to constrain the nature of the BNS remnant. All such studies have yielded non-detections, translating to upper limits on the kinetic energy scales of $\lesssim 10^{53}-10^{54}$~erg \citep{MetzgerBower2014,Horesh2016,Fong2016,Klose2019,Liu2020}. A few of these studies were based on simpler analytical models \citep{NakarPiran2011} that break down in the low-density, low ejecta mass regime in which relativistic effects are increasingly important. Recent modeling developments which have incorporated relativistic effects and the ``deep-Newtonian'' regime, coupled with deeper observations, have placed constraints on the energy imparted from the remnant of $\lesssim 10^{52}$~erg. Some previous studies have concentrated on those with anomalous X-ray emission, X-ray ``plateaus'', or extended emission \citep{MetzgerBower2014,Fong2016}, as these have been attributed to the formation of magnetars \citep{Bucciantini2012}. Others have focused on short GRBs with candidate kilonovae \citep{Horesh2016} or radio continuum surveys of short GRBs to look for optically obscured star formation \citep{Klose2019}. 

Here, we take a different approach from previous studies, and target low-redshift short GRBs ($z \lesssim 0.5$), agnostic to their X-ray behavior or association to kilonovae. Assuming that these events are associated with BNS mergers, they provide the deepest constraints on the fate of the remnant that can be attained from the short GRB population. In Section~\ref{sec:obs}, we present the details of our observations of nine short GRBs and introduce additional literature data at $1-6$~GHz. In Section~\ref{sec:modeling} we introduce our new light curve model for highly-energetic kilonova ejecta. In Section~\ref{sec:kineticenergy} we use our radio limits and the literature data, along with our new light curve modeling, to determine the allowed ejecta kinetic energies $E_{\rm ej}$ from short GRBs. In Section~\ref{sec:mtov} we compare the maximum kinetic energies $E_{\rm ej,max}$ to theoretical expectations to place constraints on $M_{\rm TOV}$, and explore the role of high-mass, GW190425-like mergers. In Section~\ref{sec:obsprospects} we explore future observational prospects in constraining BNS remnants, with a focus on observations of cosmological short GRBs, follow up of GW events, and searches in untargeted radio surveys. We end with a summary and our conclusions in Section~\ref{sec:summaryconclusion}. In this paper, we employ a standard cosmology of $H_{0}$ = 69.6 ${\rm km \, s}^{-1} \, {\rm Mpc}^{-1}$, $\Omega_{M}$ = 0.286, $\Omega_{\rm vac}$ = 0.714 \citep{Bennett2014}.

\section{Observations}
\label{sec:obs}

\tabletypesize{\normalsize}
\begin{deluxetable*}{lccccc}
\tablecolumns{6}
\tablewidth{0pc}
\tablecaption{Log of 6.0~GHz VLA Observations of Short GRBs 
\label{tab:obs}}
\tablehead {
\colhead {GRB}                &
\colhead {$z$}          &
\colhead {UT Date}           &
\colhead {$\delta t_{\rm rest}$}          &
\colhead {$F_{\nu}$} & 
\colhead {$L_{\nu}$} \\
\colhead {}                    &
\colhead {}                 &
\colhead {}              &
\colhead {(yr)}           &
\colhead {($\mu$Jy)}  &   
\colhead {(erg s$^{-1}$ Hz$^{-1}$)}   
}
\startdata
GRB\,050509B  &  0.225  &  2019 Jan 10.592  &   13.683 & $< 7.8$ & $<1.2 \times 10^{ 28 }$ \\
GRB\,060502B  &  0.287  &  2019 Jan 11.942  &  12.705 & $< 6.6$ & $<1.7 \times 10^{ 28 }$ \\
GRB\,100206A  &  0.41  &  2019 Jan 4.988  &   8.598 & $< 8.1$ & $<4.9 \times 10^{ 28 }$  \\
GRB\,130603B  &  0.356  &  2019 Jan 12.587  &   5.614 & $< 12.3$ & $<5.4 \times 10^{ 28 }$  \\
GRB\,130822A  &  0.154  &  2019 Jan 6.941  &   5.379 & $< 8.7$ & $<5.7 \times 10^{ 27 }$  \\
GRB\,140903A  &  0.351  &  2019 Feb 5.391  &   4.427 & $< 19.5$ & $<8.3 \times 10^{ 28 }$ \\
GRB\,150120A  &  0.46  &  2019 Jan 14.063  &  3.986 & $< 6.3$ & $<5.1 \times 10^{ 28 }$ \\
GRB\,150424A  &  0.3$^{a}$  &  2019 Feb 5.308  &   3.789 & $< 9.0$ & $<2.6 \times 10^{ 28 }$  \\
GRB\,160821B  &  0.16  &  2019 Jan 8.021  &   2.382 & $< 8.1$ & $<5.8 \times 10^{ 27 }$  
\enddata
\tablecomments{Upper limits correspond to $3\sigma$ confidence. \\
$^{a}$ The redshift of $z=0.3$ quoted for GRB\,150424A is based on an association to a bright, nearby galaxy \citep{gcn17758}. However it is possible that the burst is instead associated with a fainter galaxy at $z\approx 1$ \citep{Knust2017,Jin2018}. For this paper, we assume $z=0.3$. \\
{\bf References for redshifts:} (1) \citealt{Bloom2006}, (2) \citealt{Gehrels2005}, (3) \citealt{Bloom2007}, (4) \citealt{Perley2012}, (5)  \citealt{Cucchiara2013}, (6) \citealt{deUgartePostigo2014}, (7) \citealt{gcn15178}, (8) \citealt{Troja2016}, (9) \citealt{gcn17358}, (10) \citealt{gcn17758}, (11) \citealt{Lamb2019}.
 }
\end{deluxetable*}

\subsection{Sample}

Our sample comprises nine low-redshift short GRBs discovered by the {\it Swift}/Burst Alert Telescope (BAT) \citep{Gehrels2004}. The redshifts, determined from the spectroscopic redshifts of their host galaxies, are $z \approx 0.16-0.46$ (Table~\ref{tab:obs}; \citealt{Bloom2006,Gehrels2005,Bloom2007,Perley2012,Cucchiara2013,deUgartePostigo2014,gcn15178,Troja2016,gcn17358,gcn17758,Lamb2019}). This sample represents most of the known {\it Swift} short GRBs discovered in 2005-2016 with $z \lesssim 0.45$ and sky locations observable with the VLA. Based on their BAT $\gamma$-ray light curves, the durations of eight of the bursts are $T_{90}=0.024-1.20$~s (15-350~keV) while GRB\,150424A is classified as a short GRB with extended emission \citep{gcn17759,Lien2016}, resulting in a measured $T_{90}=81.0$~s. Comparing their $\gamma$-ray properties to the {\it Swift} GRB population, all of these events are classified as short-duration, spectrally-hard GRBs.

Four of the events in the sample have sub-arcsecond localization from the detection of optical afterglows (GRBs\,130603B, 140903A, 150424A, and 160821B) while the remaining five events have positional uncertainties of $\sim2''$ from the detection of their X-ray afterglows \citep{Evans2009}. Two events, GRBs\,130603B and 160821B, have detected kilonova counterparts based on their multi-band photometry, with inferred ejecta masses of $0.03-0.08\,M_{\odot}$ and $0.011\,M_{\odot}$, respectively \citep{Berger2013,Tanvir2013,Lamb2019}.

\subsection{VLA Observations}

We observed the positions of nine short GRBs with the Karl G. Jansky Very Large Array (VLA). Observations took place between 2019 January 4 and 2019 February 5 UT in either C-configuration or the hybrid CnB configuration (PI:~Fong, Program 18B-168). Each target was observed for two hours at a mean frequency of 6.0 GHz (with lower side-bands and upper side-bands centered at 4.9 GHz and 7.0 GHz, respectively). We used the Common Astronomy Software Application (CASA) pipeline products for data calibration and analysis \citep{CASA}, using 3C147, 3C286, and 3C48\footnote{Since January 2018, 3C48 has undergone flaring which may affect the flux calibration at a level of $\lesssim 5\%$. We use 3C48 only for 1 event, GRB\,060502B, and expect the effect to be negligible.} for flux calibration, and standard sources in the VLA calibrator catalog for gain calibration. We used CASA/{\tt tclean} to image the sources, employing Briggs weighting with a robust parameter of 0.5. The average beam size of the observations is $4.6'' \times 2.9''$. The details of the observations are listed in Table~\ref{tab:obs}. Other than for GRB\,100206A (described below), we do not detect any radio sources in or around the GRB positions.

At the position of GRB\,140903A, there is severe contamination by the side-lobes of an unrelated 11.3\,mJy source, NVSS\,155207+273501 \citep{Condon1998},  within the pointing field-of-view. To mitigate the effects of the bright source, the field of GRB\,140903A was calibrated and imaged outside of the standard NRAO pipeline in order to apply a peeling algorithm \citep{Noordam2004,Intema2009} to reduce the sidelobes of the source. The data were reduced in the CASA framework using standard calibrations and automatic RFI flagging with the \texttt{aoflagger} program \citep{Offringa2010,Offringa2012}. After an initial round of calibration, the bright source was subtracted using the ``rubbl-rxpackage peel'' workflow described in \citet{Williams2019}. The peeled visibilities were then inverted using multi-frequency synthesis \citep{Sault1994} to create an image of 1201 pix$^{2}$, each pixel $0.5''$ on a side. After removing the contaminating effects of this source, we do not detect a radio source at the position of GRB\,140903A.

To obtain the RMS ($\sigma_{\rm RMS}$) of each image, we use the {\tt imtool} package within {\tt pwkit} on source-free regions around the position of each GRB \citep{Pwkit}. We find $\sigma_{\rm RMS} = 2.1-6.5\,\mu$Jy, resulting in $3\sigma$ upper limits on the flux density of $F_{\nu}\lesssim 6.3-19.5\,\mu$Jy.

For GRB\,100206A, we detect a source at RA=\ra{3}{08}{39.163}, Dec=\dec{+13}{09}{29.18} on the outskirts of the XRT position (90\% confidence; \citealt{Evans2009}). This position is coincident with the centroid of the host galaxy, which is classified as a luminous infrared galaxy (LIRG; \citealt{Perley2012}). Employing a point-source model with {\tt imtool}, we measure $F_{\nu}=60.4\pm5.0\,\mu$Jy. We note that the radio flux measured is consistent with observations taken 5~years prior \citep{Klose2019}, and we attribute this emission to star formation in the host galaxy. Using standard relations between the radio flux and star formation rate (SFR; \citealt{YunCarilli2002,Perley2013}), we derive SFR=$78\pm6.5\,M_{\odot}$~yr$^{-1}$, roughly twice the SFR derived from the optical spectroscopy of $20-40\,M_{\odot}$~yr$^{-1}$ \citep{Perley2012}, indicative of obscured star formation. This result is also 1.3 times higher than the determination from radio observations of $59 \pm 10\,M_{\odot}$~yr$^{-1}$ \citep{Klose2019}, with the discrepancy due to minor differences in SFR relations used. No other radio sources are found in or near the XRT position, and we thus derive a $3\sigma$ limit of $F_{\nu} \lesssim 8.1\,\mu$Jy for GRB\,100206A.

Using the redshift of each burst, we calculate the spectral luminosity, $L_{\nu}$, as well as the rest-frame time of the observation since the {\it Swift}/BAT trigger, $\delta t_{\rm rest}$. The observations, along with model light curves (Section~\ref{sec:model_lc}) are shown in Figure~\ref{fig:lightcurve}.

\subsection{Literature Data}

\begin{figure*}[!t]
\centering
  \includegraphics[width=0.8\textwidth]{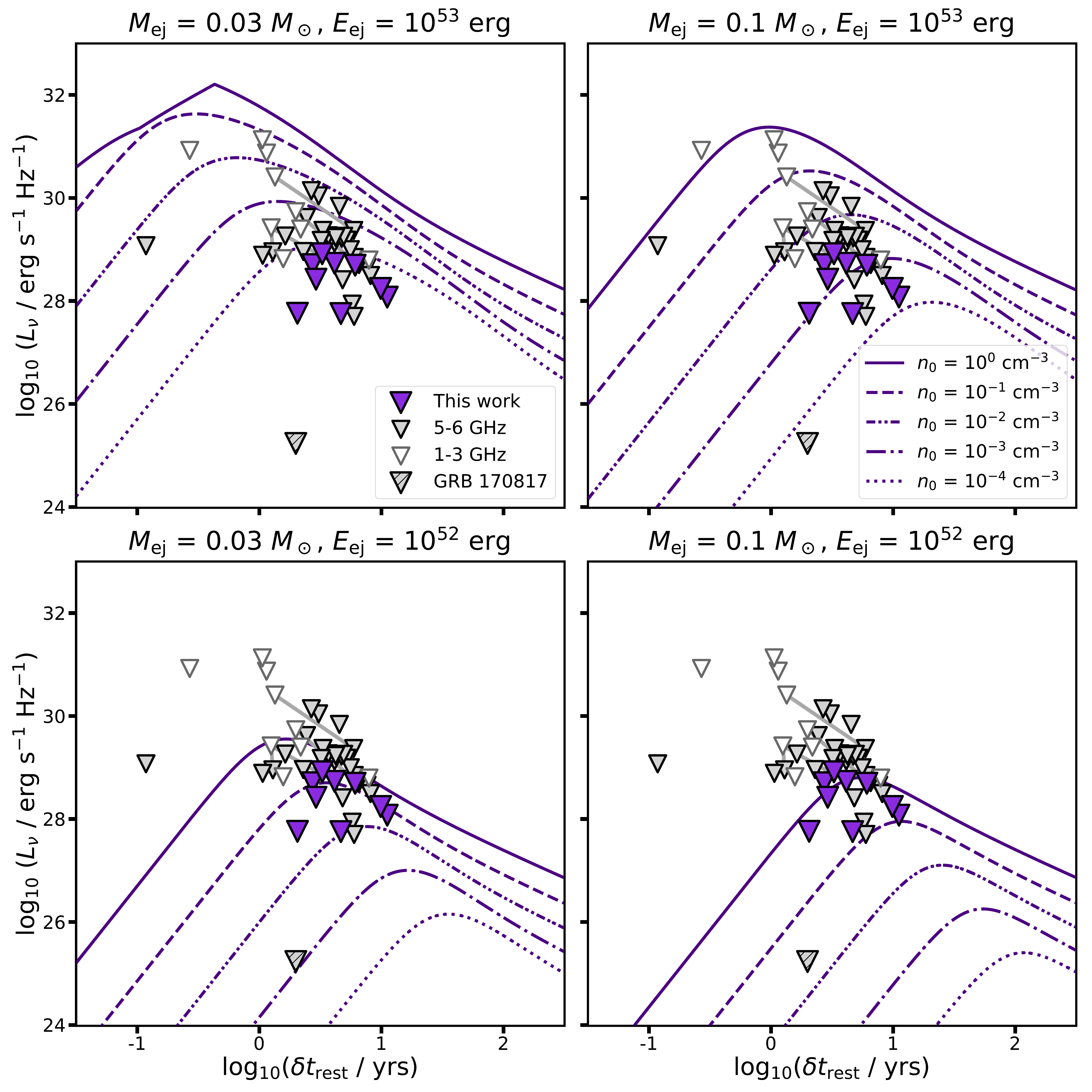}
  \vspace{-0.1in}
  \caption{6.0~GHz spectral luminosity, $L_{\nu}$ versus rest-frame time $\delta t_{\rm rest}$ of the nine low-$z$ short GRBs in our sample (purple triangles), where triangles denote $3\sigma$ upper limits. Also shown are all short GRBs with radio observations at $\delta t_{\rm rest}\gtrsim 0.1$~years taken at $5.5$~GHz and $6.0$~GHz with ATCA and the VLA (gray triangles; \citealt{Fong2016, Klose2019}) and with the VLA at $1-3$~GHz (open gray triangles; \citealt{MetzgerBower2014,Horesh2016}). Lines connect observations for the same burst. These limits are compared to 6.0~GHz light curve models computed for two sets of fixed ejecta mass $M_{\rm ej}=0.03$ and $0.1\,M_{\odot}$ and ejecta energies $E_{\rm ej}=10^{52}$ and $10^{53}$~erg at varying densities of $n_0 = 10^{-4}-1$~cm$^{-3}$, and fixed parameters $p=2.4$, $\epsilon_e = 0.1$, $\epsilon_B = 10^{-2}$. We note that a direct comparison to the 1-3~GHz limits requires an increase to the models by a small correction factor of $\lesssim 3$ ($\propto \nu^{-(p-1)/2}$). However, we show all existing radio limits here for completeness. Also shown is the latest published $6.0$~GHz limit for GW170817 (hatched triangle; \citealt{Hajela2019}). }
  \label{fig:lightcurve}
\end{figure*}

To supplement our low-redshift sample, we collect all available radio limits following short GRBs on $\delta t_{\rm rest}\gtrsim 0.1$~year timescales from the literature. We include 9 limits at 1.4~GHz, 2.1~GHz and 3.0~GHz \citep{MetzgerBower2014,Horesh2016}, 17 limits at 5.5~GHz \citep{Klose2019} and 9 limits at 6.0~GHz from our previous work \citep{Fong2016}. For each burst, we compute the $3\sigma$ limit in flux density, and use the redshift of the burst to convert to a $3\sigma$ upper limit in $L_{\nu}$. When considering both this work and the literature sample, there are multiple observations for seven events (GRBs\,050724, 070724A,  051221A, 060502B, 100206A, 130603B, 150424A). The total number of short GRBs with deep observations on these timescales is thus 27 bursts. The literature data are also shown in Figure~\ref{fig:lightcurve}.

\section{Lightcurve Modeling}
\label{sec:modeling}

In the following, we discuss our new lightcurve modeling. Compared to the analytic framework first laid out by \citet{NakarPiran2011}, our present models incorporate the full dynamics of a single-velocity-shell ejecta including the transition between coasting and decelerating phases, relativistic dynamics, relativistic timing and Doppler effects on the lightcurve, and the deep-Newtonian regime. These factors combined have important effects on the pre-peak behavior of the light curves. A similar model was also recently applied by \cite{Liu2020} to a sample of previously-published short GRB radio limits and we provide a comparison in Section~\ref{sec:obsprospects}. 

The radio signatures of kilonovae ejecta were first discussed by \cite{NakarPiran2011} who showed that this emission typically peaks at the deceleration time, $t_{\rm dec}$, at which the ejecta dynamics transition from the coasting ejecta-dominated phase to the Sedov-Taylor phase.
\cite{NakarPiran2011} discussed the case of non-relativistic ejecta ($\Gamma \beta \lesssim 1$) relevant to the initial ejection velocities of material from BNS mergers. The deposition of additional energy into this ejecta by a long-lived magnetar remnant would accelerate this ejecta to potentially relativistic velocities. 
As this is precisely the scenario we wish to constrain, we extend the \cite{NakarPiran2011} model to account for relativistic dynamics of the ejecta and relativistic effects on the resulting lightcurve (see also \citealt{HotokezakaPiran15,Horesh2016,Liu2020}).

The dynamics of an ejecta with kinetic energy $E_{\rm ej}$ and mass $M_{\rm ej}$ depend on its initial Lorentz factor $\Gamma_0 = 1 + E_{\rm ej}/M_{\rm ej}c^2$ and corresponding velocity $\beta_0 c = c ( 1-\Gamma_0^{-2} )^{1/2}$, in combination with the ambient density $n_0$. With these parameters, the characteristic timescale (in the observer frame) at which the radio light-curve peaks is \citep{HotokezakaPiran15}
\begin{equation}
\label{eq:tdec}
    t_{\rm dec} \sim \left(\frac{3E_{\rm ej}}{4\pi m_p c^5 n_0 \Gamma_0 (\Gamma_0-1) \beta_0^3}\right)^{1/3} \frac{(1+z)}{\Gamma_0^2}
    ,
\end{equation}
where the final term $(1+z)/\Gamma_0^2$ is responsible for transforming between the blast-wave rest-frame and the observer frame, and accounts for cosmological redshift of the source.

We model the ejecta dynamics within the ``homogeneous shell approximation'' following \citet{Peer12} (see also \citealt{Huang+99,Nava+13}). This one-zone model allows us to numerically integrate the forward-shock dynamics for arbitrary $\Gamma_0$ and correctly reproduce the coasting phase at $t \ll t_{\rm dec}$ and Blandford-McKee (Sedov-Taylor) solutions in the ultra-relativistic (non-relativistic) regimes at $t \gg t_{\rm dec}$.

We calculate synchrotron emission from the shocked material assuming electrons at the shock-front are accelerated into a non-thermal population that shares a fraction $\epsilon_e$ of the shock power, and that magnetic fields are amplified behind the shock-front such that $B = \sqrt{8 \pi \epsilon_B u_{\rm th}}$ where $u_{\rm th}$ is the post-shock thermal energy density and $\epsilon_B$ a constant $<1$ (e.g. \citealt{Sari+98}).

We account for effects of the deep-Newtonian regime on the non-thermal electron distribution \citep{SironiGiannios13}, relevant when the ejecta velocity is $\lesssim 0.2 c \left( \bar{\epsilon}_e / 0.1 \right)^{-1/2}$, where $\bar{\epsilon}_e \equiv 4 \epsilon_e (p-2)/(p-1)$ and $2<p<3$ is the power-law index describing the non-thermal electron distribution, $dN_e/d\gamma \propto \gamma^{-p}$ at $\gamma \geq \gamma_m$ (where $\gamma$ is the electron Lorentz factor).  In what follows we assume a value of $\epsilon_e = 0.1$, motivated by the findings of first principles particle-in-cell numerical simulations into weakly magnetized plasma characteristic of the ISM (e.g.~\citealt{Spitkovsky08}).  In the deep-Newtonian regime, the minimal Lorentz factor of shock-accelerated electrons is $(\gamma_m-1) \lesssim 1$ and only a small fraction of electrons are accelerated to $\gamma \gtrsim 2$ where they can emit synchrotron radiation.
This has an important impact on the resulting radio light-curves. 

For most of the physically-relevant parameter space, emission at $\sim$GHz radio frequencies is in the slow-cooling optically-thin regime; however our model also fully allows for fast-cooling electrons (following \citealt{Sari+98}). Synchrotron self-absorption is taken into account in an approximate manner,\footnote{This approximation is correct to within an order-unity factor (dependant on $p$) of the full treatment of synchrotron self-absorption in the non-relativistic spherical regime and suffices for purposes of our present work (and see \citealt{Granot+99} for a relativistic treatment).} 
by limiting the spectral luminosity to a maximum defined by $L_\nu = 8\pi^2 \Gamma R^2 kT(\gamma) {\nu^\prime}^2 / c^2$ where $R$ is the blast-wave radius, $kT(\gamma) \approx \gamma m_e c^2/3$, $\gamma = \max \left[ \gamma_m, \left({2\pi m_e c \nu^\prime}/{e B}\right)^{1/2} \right]$, and $\nu^\prime$ is the emission frequency in the blast-wave rest frame, related to the observed frequency through $\nu^\prime \approx \nu \Gamma (1-\beta) (1+z)$.
We find that synchrotron self-absorption is only relevant in a small subset of cases where the initial ejecta Lorentz factor is very large (large $E_{\rm ej}$ and small $M_{\rm ej}$).

Compared to the \cite{NakarPiran2011} analytic models that were used in \cite{MetzgerBower2014}, \cite{Fong2016}, and \cite{Klose2019}, relativistic effects (for high $E_{\rm ej}$) cause the light-curve to peak earlier (by a factor of $\sim \Gamma_0^{-8/3}$; eq.~\ref{eq:tdec}) and at a larger flux. 
Relativistic effects were first accounted for by \cite{Horesh2016}, with a slightly different dynamical model than we adopt here. Their model, based on \cite{Piran+13}, produces the correct temporal scaling laws in the coasting, Blandford-McKee, and Sedov-Taylor regimes, however may differ by factors of order unity from the exact quantitative dynamics implied by these analytic solutions.
The strong dependence of the radio luminosity on blast-wave velocity ($\propto \beta^{(5p-3)/2}$ in the non-relativistic non-deep-Newtonian regime) implies that even factor of $\sim$two differences in velocity can amount to an $\gtrsim$order-of-magnitude difference in luminosity.
We have carefully verified that our model asymptotes to the exact coasting / Blandford-McKee / Sedov-Taylor solutions in the appropriate limits, including subtleties of distinguishing between ejecta (contact discontinuity) and shock velocities that are neglected in some models.

As in previous work, our model assumes spherical symmetry (though see \citealt{MargalitPiran15}) and does not account for velocity-structure of the ejecta. The latter amounts to a conservative assumption: a radial velocity profile would enhance the early ($\lesssim t_{\rm dec}$) light-curve leading to more stringent upper-limits on $E_{\rm ej}$.

Finally, we note that our model is geared towards constraining highly-energetic kilonova ejecta due to possible magnetar energy deposition. Radio emission from ``normal'' kilonovae ejecta, with $E_{\rm ej} \sim 10^{51} \, {\rm erg}$, may in fact be inhibited for a significant period of time due to interaction of the GRB jet with the ambient medium ahead of the ejecta \citep{MargalitPiran20}. Our model does not account for this effect, which is likely negligible for highly-energetic (fast) ejecta.

\section{Short GRB Remnant Kinetic Energy Constraints}
\label{sec:kineticenergy}
Here, we use the radio limits derived in Section~\ref{sec:obs} and the light curve modeling described in Section~\ref{sec:modeling} to constrain the magnetar remnant energy that is transferred to the ejecta as kinetic energy ($E_{\rm ej}$) in short GRBs.

\subsection{Lightcurve Comparisons}
\label{sec:model_lc}
We generate model light curves for a range of circumburst densities to represent those inferred from broad-band analysis of short GRB afterglows, $n_0=10^{-4}-1$~cm$^{-3}$ (Figure~\ref{fig:lightcurve}; \citealt{Fong2015}). We compute light curves for two sets of ejecta masses, $M_{\rm ej}=0.03\,M_{\odot}$ and $0.1\,M_{\odot}$, which are chosen to represent the range of inferred ejecta masses for kilonovae following short GRBs \citep{Ascenzi2019} and GW170817 ($M_{\rm ej} \approx 0.05\,M_{\odot}$; \citealt{Villar2017,Arcavi2018}). We also choose two fiducial kinetic energies deposited into the ejecta, $E_{\rm ej} = 10^{52}$~erg and $10^{53}$~erg, where the former represents a typical energy extractable from a supramassive NS (e.g., \citealt{MargalitMetzger2019}), while the latter represents the maximum energy that can be extracted from a magnetar with $M = M_{\rm TOV} \sim 2.2\,M_{\odot}$ \citep{Metzger+15}. We fix the value of $\nu = 6$~GHz, $p=2.4$, $\epsilon_e = 0.1$,  $\epsilon_B=10^{-2}$, and $\zeta_e = 1$, where $\zeta_e$ is the fraction of electrons participating in diffusive shock acceleration out of total electrons. The resulting model light curves ($L_{\nu}$ versus $\delta t_{\rm rest}$), grouped by ejecta mass and ejecta energies, are displayed in Figure~\ref{fig:lightcurve}.

As expected, we find that the radio signal is brightest for the $M_{\rm ej}=0.03\,M_{\odot}$ and $E_{\rm ej}=10^{53}$~erg case, and the predicted luminosities decrease for larger values of ejecta mass and lower ejecta energies. In addition, for a given set of fixed parameters, the peak timescales increase as density decreases; this can be intuitively explained since shocks in lower density environments take longer to sweep up a mass comparable to their own (Equation~\ref{eq:tdec}). A basic comparison to the low-redshift short GRB limits shows that in the most optimistic case, models with extremely low densities of $n_0 \gtrsim 10^{-4}$~cm$^{-3}$ are ruled out. Moreover, in the most pessimistic case ($M_{\rm ej}=0.1\,M_{\odot}$, $E_{\rm ej}=10^{52}$~erg), the low-redshift short GRB limits still provide meaningful constraints of $n_0 \lesssim 0.1-1$~cm$^{-3}$ (Figure~\ref{fig:lightcurve}). Overall, the short GRB limits presented here, in the context of new light curve modeling, are significantly more constraining than many previous works \citep{MetzgerBower2014,Fong2016,Klose2019,Liu2020}.

\tabletypesize{\normalsize}
\begin{deluxetable*}{lcccccc}
\tablecolumns{7}
\tablewidth{0pc}
\tablecaption{Short GRB Afterglow and Neutron Star Remnant Properties
\label{tab:param}}
\tablehead {
\colhead {GRB}                &
\colhead{z}                 &
\colhead {$n_0$}           &
\colhead{$p$}             &
\colhead{$\varepsilon_B$}       &
\colhead {$E_{\rm ej, max}$ (0.03$\,M_{\odot}$)}           &
\colhead {$E_{\rm ej, max}$ (0.1$\,M_{\odot}$)}            \\
\colhead {}                    &
\colhead {}                    &
\colhead {(cm$^{-3}$)}                 &
\colhead {}                    &
\colhead {}                    &
\colhead {($10^{52}$~erg)}              &
\colhead {($10^{52}$~erg)}                                   
}
\startdata
\multicolumn{7}{c}{\emph{This work}} \\
\hline
GRB\,050509B & 0.225 & $<$0.55 & 2.52 & $10^{-2}$ & 0.73-17.58$^a$ & 0.79-39.66$^a$\\
GRB\,060502B & 0.287 & 0.01 & 2.4 & $10^{-2}$ & 1.59 & 2.98\\
GRB\,100206A & 0.41 & 0.01 & 2.4 & $10^{-2}$ & 2.06 & 4.64\\
GRB\,130603B & 0.356 & $0.31^{+0.08}_{-0.04}$ & 2.7 & $10^{-2}$ & 1.1 & 1.98\\
GRB\,130822A & 0.154 & 0.01 & 2.4 & $10^{-2}$ & 1.02 & 3.09\\
GRB\,140903A & 0.351 & $3.40^{+2.9}_{-1.6}\times 10^{-3}$ & 2.27 & $10^{-3}$ & 6.72 & 14.08\\
GRB\,150120A & 0.46 & 0.01 & 2.4 & $10^{-2}$ & 2.13 & 6.24\\
GRB\,150424A & 0.3 & $1.98^{+4.4}_{-1.4}\times 10^{-4}$ & 2.4 & $10^{-2}$ & 5.58$^b$ & 17.58$^b$\\
GRB\,160821B & 0.16 & $0.13^{+0.05}_{-0.04}$ & 2.36 & $10^{-2}$ & 0.56 & 1.77\\
\hline
\multicolumn{7}{c}{\emph{Fong et al. 2016}} \\
\hline
GRB\,050724A & 0.257 & $0.89^{+0.58}_{-0.49}$ & 2.29 & $10^{-4}$ & 5.79 & 6.01\\
GRB\,051221A & 0.546 & $0.14^{+0.05}_{-0.04}$ & 2.24 & $10^{-2}$ & 2.06 & 2.87\\
GRB\,070724A & 0.457 & $9.30^{+210.0}_{-9.2}\times 10^{-5}$ & 2.24 & $10^{-2}$ & 11.28 & 25.45\\
GRB\,080905A & 0.122 & $7.10^{+610.0}_{-7.1}\times 10^{-4}$ & 2.06 & $10^{-2}$ & 2.87 & 8.08\\
GRB\,090510 & 0.903 & $6.40^{+100.0}_{-6.0}\times 10^{-5}$ & 2.65 & $10^{-2}$ & 25.45 & 47.72\\
GRB\,090515 & 0.403 & 0.01 & 2.4 & $10^{-2}$ & 2.87 & 6.72\\
GRB\,100117A & 0.915 & $1.2^{+0.9}_{-0.3}$ & 2.36 & $10^{-2}$ & 2.06 & 2.98\\
GRB\,101219A & 0.718 & $2.40^{+97.0}_{-2.3}\times 10^{-4}$ & 2.73 & $10^{-2}$ & 11.7 & 28.43\\
\hline
\multicolumn{7}{c}{\emph{Klose et al. 2019}} \\
\hline
GRB\,061006 & 0.438 & $1.20^{+29.0}_{-1.1}\times 10^{-4}$ & 2.39 & $10^{-2}$ & 13.08 & 27.4\\
GRB\,061201 & 0.111 & $2.70^{+120.0}_{-2.6}\times 10^{-4}$ & 2.35 & $10^{-2}$ & 3.33 & 9.73\\
GRB\,061210A & 0.41 & 0.01 & 2.4 & $10^{-2}$ & 3.45 & 6.48\\
GRB\,070729 & 0.8 & 0.01 & 2.4 & $10^{-2}$ & 8.39 & 12.6\\
GRB\,070809 & 0.219 & $1.20^{+30.0}_{-1.1}\times 10^{-4}$ & 2.12 & $10^{-2}$ & 7.23 & 18.93\\
GRB\,080123 & 0.496 & 0.01 & 2.4 & $10^{-2}$ & 3.86 & 8.39\\
GRB\,090621B & 0.5$^c$ & $1.0^{+0.52}_{-0.27}$ & 2.64 & $10^{-2}$ & 2.06 & 2.39\\
GRB\,100816A & 0.805 & 0.01 & 2.4 & $10^{-2}$ & 2.87 & 8.39\\
GRB\,101224A & 0.454 & 0.01 & 2.4 & $10^{-2}$ & 3.45 & 8.39\\
GRB\,130515A & 0.5$^c$ & 0.01 & 2.4 & $10^{-2}$ & 3.86 & 11.28 
\enddata
\tablecomments{Upper limits, $E_{\rm ej,max}$ calculated for two ejecta masses, $M_{\rm ej} = 0.03\,M_{\odot}$ and $M_{\rm ej}=0.1\,M_{\odot}$ at the median inferred density inferred from the afterglow. If afterglow constraints do not exist, we use fiducial values, $p=2.4$, $n=0.01$~cm$^{-3}$, $\epsilon_e=0.1$, and $\epsilon_B=10^{-2}$. \\
$^{a}$ Range is quoted corresponding to $n_0 = 10^{-6}-0.55$~cm$^{-3}$ where the upper bound is set by the afterglow parameters. \\
$^{b}$ We note that if we assume $z=1$ for GRB\,150424A, the values for $E_{\rm ej, max}=11.70$ (for $0.03\,M_{\odot}$) and $E_{\rm ej, max}=31.77$ (for $0.1\,M_{\odot}$).
\\
$^{c}$ Redshift is set to a fiducial value of 0.5 for GRBs with no known redshift}
\end{deluxetable*}

\subsection{Afterglow Parameter Constraints}
\label{sec:ag}
Since there are degeneracies between model parameters, we first collect available constraints on the values of $p$, $\epsilon_B$, and $n_0$ from existing broad-band afterglow fitting. For the low-redshift sample presented here, only two GRBs have previously determined values, GRB\,130603B and GRB\,140903A. For GRB\,130603B: $n_0=0.31^{+0.08}_{-0.04}$~cm$^{-3}$ and $p=2.70$ ($\epsilon_e = 0.1$, $\epsilon_B = 10^{-2}$) and for GRB\,140903A: $n_0=3.40^{+2.9}_{-1.6} \times 10^{-3}$~cm$^{-3}$ and $p=2.27$ $(\epsilon_e = 0.1$,  $\epsilon_B = 10^{-3}$; \citealt{Fong2015}).

For GRBs\,050509B, 150424A, 160821B, we use available afterglow data at early times to place constraints on these parameters. We use the standard synchrotron model for a relativistic blast-wave \citep{GranotSari2002}, which provides a mapping of the afterglow spectral and temporal evolution to the isotropic-equivalent afterglow kinetic energy ($E_{\rm K,iso}$), $n_0$, $\epsilon_e$ and $\epsilon_B$. In general, we choose fiducial values for the microphysical parameters. In particular, we fix $\epsilon_e=0.1$, motivated both observationally with GRB afterglow fitting which finds $\epsilon_e \sim 0.1$ (\citealt{Panaitescu2001}, \citealt{Ryan2015}, \citealt{Beniamini2017}), and theoretically by first-principles particle-in-cell simulations, which find $\sim 10 \%$ of the kinetic energy from shocks can be deposited into non-thermal particles \citep{Spitkovsky08}. For $\epsilon_B$, short GRBs with broad-band afterglow detections (radio, optical and X-rays) tend to favor values $\epsilon_B<0.1$ for fixed $\epsilon_e=0.1$, although models with $\epsilon_B<10^{-3}$ are routinely found to provide inadequate matches to the broad-band data \citep{Fong2015}. Thus, we employ a fiducial value of $\epsilon_B=10^{-2}$ unless otherwise stated. Finally, we choose a fiducial value of $p = 2.4$, motivated by the median value found for short GRBs \citep{Fong2015}.

For GRB\,050509B, we use the X-ray afterglow detection ({\it Swift}/XRT; \citealt{050509BXray}), as well as the optical  \citep{050509Boptic} and radio upper limits \citep{050509Bradio} to determine a value of $p=2.52$ and a limit on the circumburst density of $n_0 \lesssim 0.55$~cm$^{-3}$ for $\epsilon_e=0.1$, $\epsilon_B=10^{-2}$ (Table~\ref{tab:param}). For GRB\,150424A, we use the X-ray \citep{Evans2009}, optical \citep{150424Auvot} and 9.8~GHz VLA (Program 15A-235; updated from \citealt{150424Aradio}) afterglow detections to determine a value of $p = 2.40$ and a low density of $n_0 =  1.98^{+4.4}_{-1.4} \times 10^{-4}$~cm$^{-3}$ (assuming $\epsilon_e=0.1$, $\epsilon_B=10^{-2}$). Finally, for GRB\,160821B, we use the X-ray \citep{Evans2009}, optical \citep{Lamb2019} and 5.0~GHz VLA detection (Program 15A-235; updated from \citealt{GCN19854}) to determine $p=2.36$ and $n_0=0.13^{+0.05}_{-0.04}$~cm$^{-3}$ (assuming $\epsilon_e = 0.1$, $\epsilon_B=10^{-2}$; Table~\ref{tab:param}).

For the remaining short GRBs in the low-redshift sample, the afterglow data are too sparse to adequately constrain these parameters. Thus, we assume fiducial values for these bursts (Table~\ref{tab:param} and Section~\ref{sec:EKlimits}). We also collect measurements or constraints on the value of $p$ and $n_0$ for 18 additional short GRBs with radio limits at $5.5$ and $6.0$~GHz \citep{Fong2016,Klose2019} from \citet{Fong2015}. The afterglow parameters for all short GRBs used in this work are summarized in Table~\ref{tab:param}.

\begin{figure*}
\centering
  \includegraphics[width=0.9\textwidth]{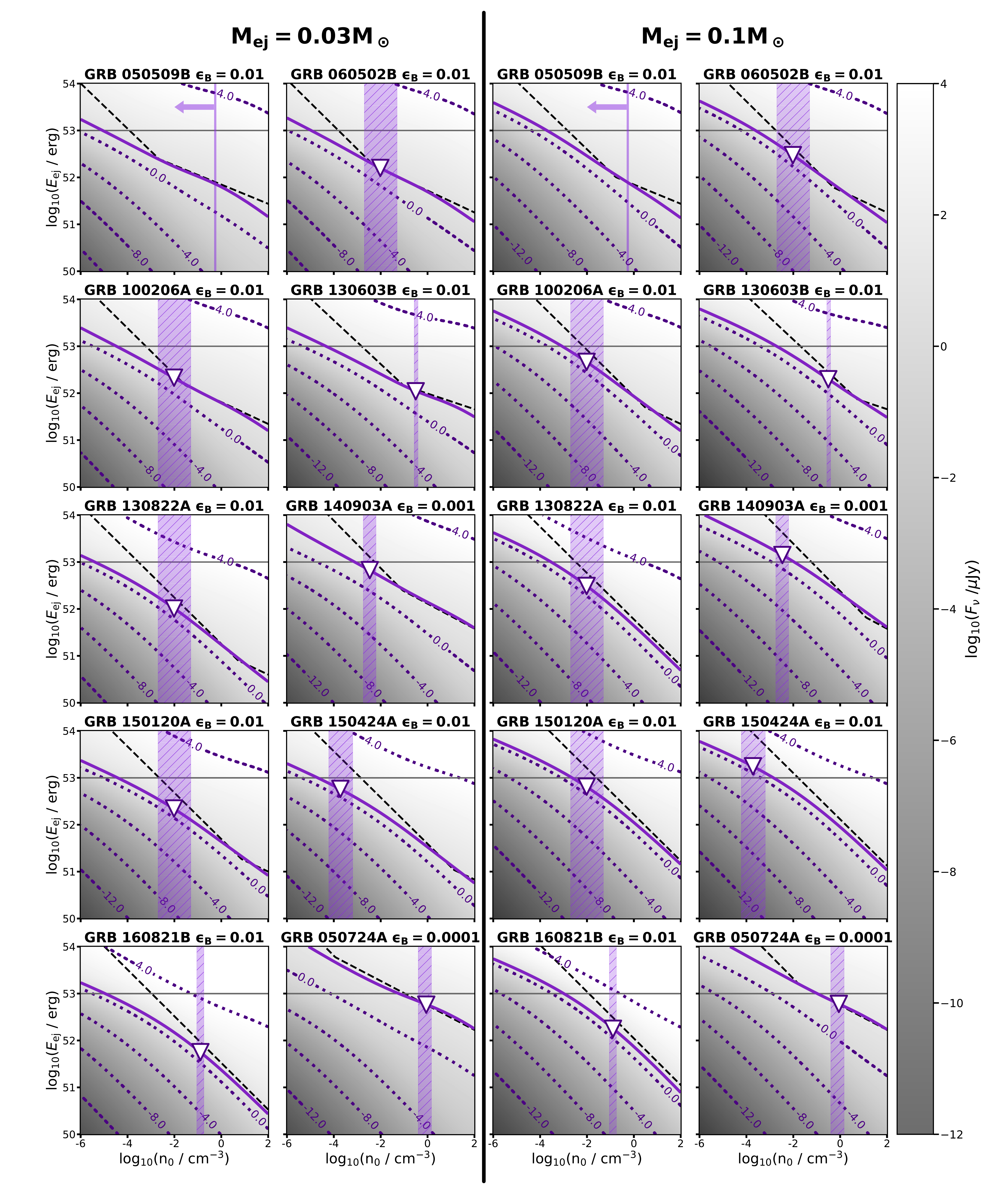}
  \vspace{-0.0in}
  \caption{Ejecta energy $E_{\rm ej}$ versus circumburst density $n_0$ parameter space for ten short GRBs (nine in this work, and GRB\,050724A from \citealt{Fong2016}) for two ejecta masses, $M_{\rm ej} = 0.03 M_\odot$ (left) and $M_{\rm ej} = 0.1 M_\odot$ (right). The solution for each GRB is shown (purple solid line), in which parameter space above the line is ruled out and parameter space below the line is allowed. Also shown is the flux density solution for each GRB at the rest-frame time of the observation (gray-scale map), and contours (indigo dotted lines) that are spaced by $10^4\,\mu$Jy. The purple shaded region corresponds to the density constraints used from afterglow measurements, with a fiducial range of $n_0 = (0.2-5) \times 10^{-2}$~cm$^{-3}$ if no constraint is available. The triangle in each frame is the upper limit placed on $E_{\rm ej}$ at the median density. A comparison to the analytical model solution (black dashed line; \citealt{NakarPiran2011}) demonstrates that the largest deviations are for high values of $E_{\rm ej}$ and low values of $n_0$. A gray horizontal line at $10^{53}$~erg denotes the maximum extractable energy expected for a stable magnetar with $M\sim 2.2 M_\odot$.}
  \label{fig:Evsnmej0.03}
\end{figure*}


\subsection{Inferred Limits on Ejecta Kinetic Energies}
\label{sec:EKlimits}

We next compare the late-time radio observations to a combination of the deposited energy $E_{\rm ej}$ and circumburst density $n_0$, by determining the parameter space ruled out by each 6.0~GHz limit. For each short GRB in the low-redshift sample, we generate a grid of $62,500$ models with the appropriate values of $p$ and $\epsilon_B$ inferred from the broad-band afterglow, with $\epsilon_e = 0.1$ and $\zeta_e = 1$ (Table~\ref{tab:param}). For bursts which lack available information, we assume fiducial values of $p=2.4$ and $\epsilon_B=10^{-2}$. We also re-analyze GRB\,050724A first presented in \citet{Fong2016}, using the appropriate values for $\epsilon_B$ and $n_0$.

The fixed $E_{\rm ej}-n_0$ grid is defined by 250 log-spaced points in each parameter, with the ranges $E_{\rm ej}=10^{50}-10^{54}$~erg and $n_0 = 10^{-6}-10^2$~cm$^{-3}$. The minimum and maximum circumburst densities are chosen to represent the extremes in which short GRBs are expected to occur, from a low value for the intergalactic medium (IGM) particle densities to those of star-forming regions. We note that the median density of short GRBs, commensurate with their kicked progenitors, is $n_0\approx 10^{-3}-10^{-2}$~cm$^{-3}$ \citep{Fong2015}.

To translate these models to a $E_{\rm ej}-n_0$ parameter space, for each model we calculate the value of $F_{\nu}$ at the rest-frame time of observation ($\delta t_{\rm rest}$). This flux mapping is displayed as a gray-scale gradient in Figure~\ref{fig:Evsnmej0.03}. We also determine the corresponding solution (solid purple line) represented by the measured limit, where the parameter space below each limit is allowed while parameter space above each limit is ruled out (Figure~\ref{fig:Evsnmej0.03}). Notably, most of the solutions corresponding to the low-redshift events can rule out parameter space below the $10^{53}$~erg maximum energy for a wide range of densities. For a direct comparison, we derive the same solution using the analytical models from \citet{NakarPiran2011}, shown as a black dashed line. As expected, our models deviate significantly from the analytical solutions in the high-energy, low-density parameter space when relativistic effects are expected to be strongest (see also, \citealt{Liu2020}).

We determine the upper limit on the energy for each burst ($E_{\rm ej, max}$), by imposing a density prior and fixing the densities to those derived from the broad-band afterglow (Section~\ref{sec:ag}; Table~\ref{tab:param}). 

\begin{figure*}
\centering
  \includegraphics[width=0.9\textwidth,trim={0cm 0cm -1cm 0cm}]{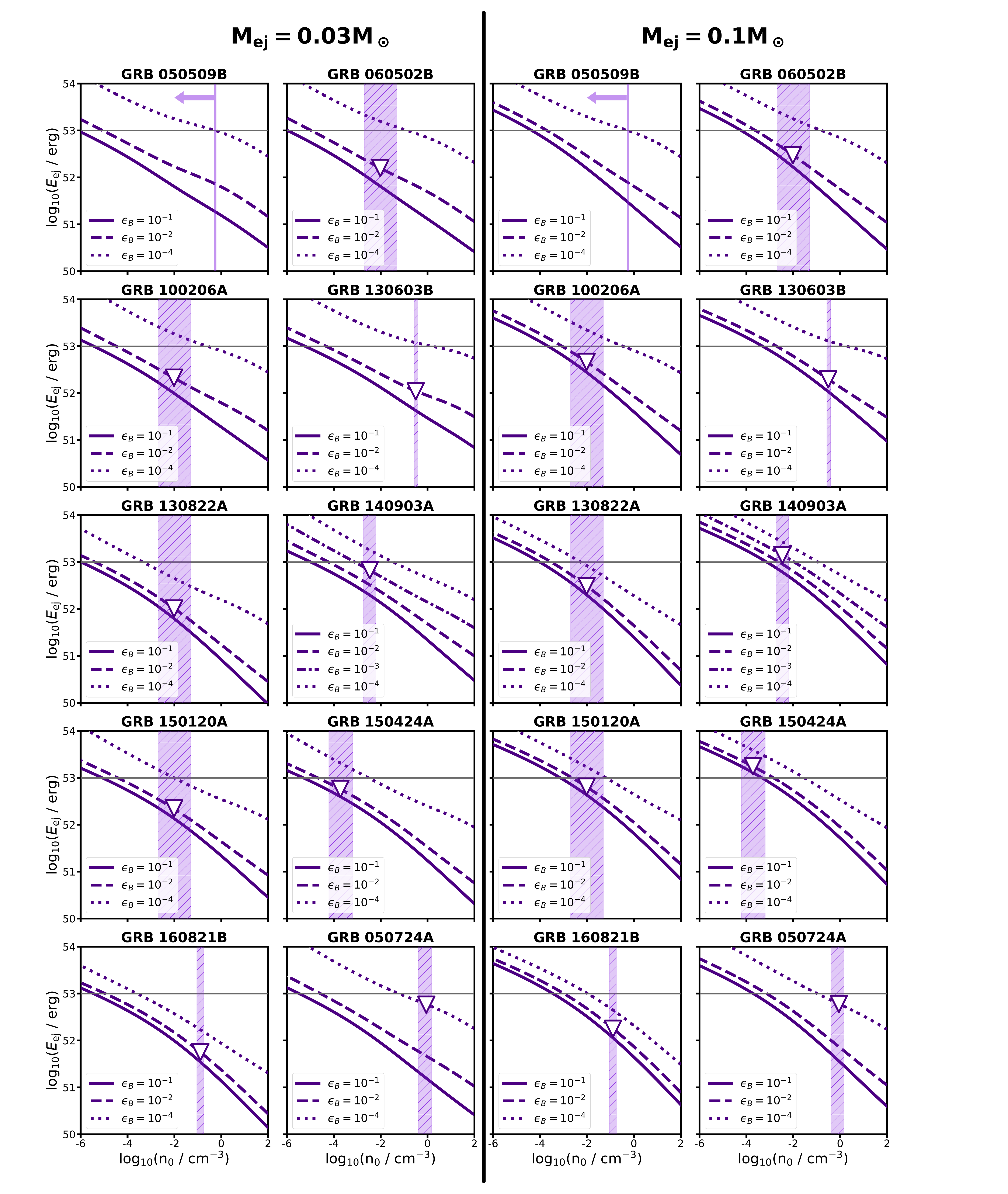}
  \vspace{-0.0in}
  \caption{Ejecta energy $E_{\rm ej}$ versus circumburst density $n_0$ parameter space for ten short GRBs for two ejecta masses, $M_{\rm ej} = 0.03 M_\odot$ (left) and $M_{\rm ej} = 0.1 M_\odot$ (right). Curves represent limits for three values of $\epsilon_B = 10^{-1}$, $10^{-2}$ (fiducial), and $10^{-4}$, except in the case of GRB\,140903A in which the additional line represents the best-fit $\epsilon_B = 10^{-3}$. The purple shaded regions correspond to the density constraints from the afterglow, or the fiducial range if no constraints were found. Triangles correspond to the value of $E_{\rm ej,max}$ of the median density inferred for the fiducial value of $\epsilon_B=10^{-2}$, except in the cases of GRB\,140903A ($\epsilon_B=10^{-3}$) and GRB\,050724A ($\epsilon_B = 10^{-4} $). A gray horizontal line at $10^{53}$~erg denotes the maximum extractable energy expected for a stable magnetar with $M\sim 2.2 M_\odot$.}
  \label{fig:Evsnmej_epsB}
\end{figure*}

The values of $E_{\rm ej,max}$ for the ten bursts for both ejecta mass scenarios are shown in Figures~\ref{fig:Evsnmej0.03} and also listed in Table~\ref{tab:param}. Finally, motivated by the fact that we do not know the true value of $\epsilon_B$, and this in turn has an effect on the inferred maximum energies, we calculate solutions in the $E_{\rm ej}-n_0$ parameter space for $\epsilon_B=10^{-1}$ and $\epsilon_B=10^{-4}$ (see \citealt{Liu2020} for an in-depth exploration of the dependence on $\epsilon_B$). Decreasing $\epsilon_B$ overall weakens the constraints on the parameter space (Figure~\ref{fig:Evsnmej_epsB}). However, a lower value of $\epsilon_B$ also results in an increase in the inferred circumburst density \citep{GranotSari2002}. Conversely, increasing $\epsilon_B$ overall strengthens the constraints on the parameter space and results in a decrease in the inferred circumburst density. The end result is a slight decrease in the values of $E_{\rm ej, max}$: for instance, for GRB\,130603B a factor of 10 increase from $\epsilon_B=10^{-2}$ to $\epsilon_B=10^{-1}$ results in a decrease in the ejecta energy limit by a factor of $\approx 1.7$ ($E_{\rm ej,max}= 1.10\times 10^{52}$~erg to $0.65 \times 10^{52}$~erg for $M_{\rm ej}=0.03\,M_{\odot}$).

\begin{figure}[t]
\centering
  \includegraphics[width=0.47\textwidth]{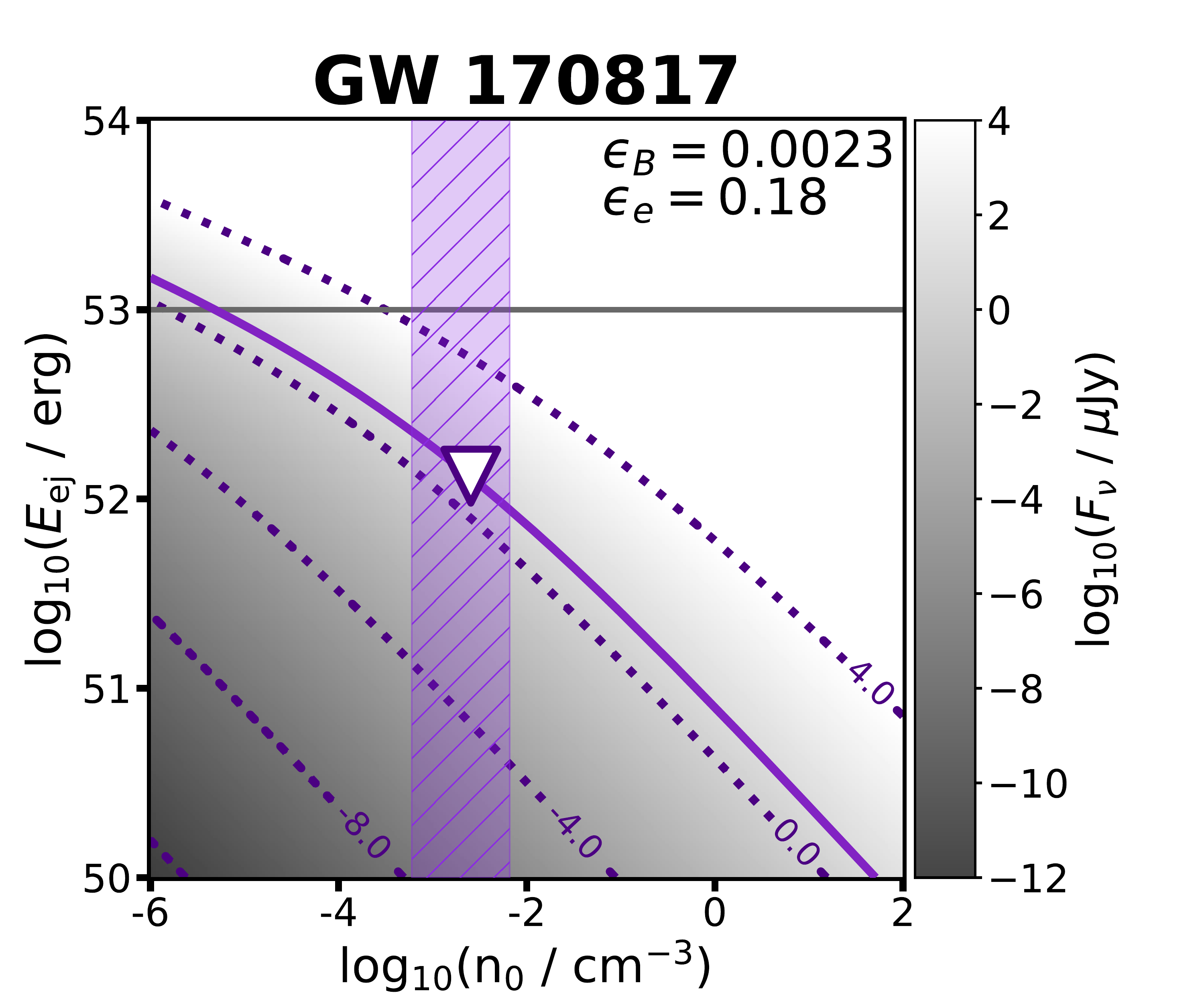}
    \vspace{-0.13in}
  \caption{Ejecta energy $E_{\rm ej}$ versus circumburst density $n_0$ parameter space for GW\,170817 for ejecta mass $M_{\rm ej} = 0.05 M_\odot$ and microphysical parameters $\epsilon_e = 0.18$ and $\epsilon_B = 0.0023$, found by \citealt{Hajela2019}. The solution for GW\,170817 is shown (purple solid line), in which parameter space above the line is ruled out and parameter space below the line is allowed. Also shown is the flux density solution for GW\,170817 at the rest-frame time of the observation (gray-scale map), and contours (indigo dotted lines) spaced by $10^4\,\mu$Jy. The purple shaded region corresponds to the density constraints used from afterglow measurements.  A gray horizontal line at $10^{53}$~erg denotes the maximum extractable energy expected for a stable magnetar with $M\sim 2.2 M_\odot$.}
  \label{fig:GW170817}
\end{figure}

To constrain the fastest ejecta deposited by a magnetar remnant in the BNS merger GW\,170817 \citep{gw170817}, we perform the same analysis to create an $E_{\rm}-n_0$ parameter space. We use the median of the physical parameters determined by \citet{Hajela2019} of $p=2.15$, $\epsilon_e = 0.18$, and $\epsilon_B = 0.0023$ (see also: \citealt{Makhathini2020}). We set the ejecta mass to be $M_{\rm} = 0.05 M_\odot$ \citep{Villar2017,Arcavi2018}. This flux mapping is displayed as gray-scaled gradient in Figure~\ref{fig:GW170817}. We also determine the corresponding solution (solid purple line) represented by the measured $3\sigma$ limit at $\delta t_{\rm rest} \approx2$~years of $F_\nu < 8.4\mu$Jy \citep{Hajela2019}. We determine the upper limit on the energy  for GW\,170817 by imposing a density prior and fixing the density to $2.5\times10^{-3}\rm~cm^{-3}$. We find a value of $E_{\rm ej, max} = 1.32 \times 10^{52}$~erg (Figure~\ref{fig:GW170817}). 

Finally, we compute the probability distribution function of ejecta energies for each burst, with and without the circumburst density prior. We note that without the density prior, the shape of each distribution is mainly governed by our choice of density grid and the flux upper limit, overall resulting in lower values of $E_{\rm ej, max}$ (Figure~\ref{fig:epdf}). We show both distributions for completeness, as we assumed a fiducial density range for $\sim1/3$ of the sample; however, the results following are based on use of the density priors. In this analysis, we include 18 additional bursts with $5.5$ and $6.0$~GHz observations from the literature \citep{Klose2019,Fong2016}, representing the full available sample of meaningful observations at these frequencies.We use spectroscopic redshifts for all events except for two for which we assume $z=0.5$ (the median of the short GRB population; \citealt{Fong2017}). We present a revised redshift for GRB\,101224A of $z=0.454$ based on spectroscopy obtained with the Large Binocular Telescope (LBT; PI Fong), which we use in this analysis. We combine the PDFs and normalize the area under the PDF to unity, to create cumulative distribution functions, shown in Figure~\ref{fig:epdf}.

\section{Short GRB Constraints on \texorpdfstring{$M_{\rm TOV}$}{Lg}}
\label{sec:mtov}
\subsection{Predictions for Energy Output for varying \texorpdfstring{$M_{\rm TOV}$}{Lg}}
We now compare the constraints on the maximum energy deposited from NS remnants to theoretical expectations. The characteristic energy scale of kilonovae ejecta without additional magnetar energy deposition is $E_{\rm ej} \lesssim 10^{50.5} \, {\rm erg}$. The rotational energy of a post-merger remnant magnetar can reach $\sim 10^{53} \, {\rm erg}$. If a NS remnant remains stable for sufficiently long timescales following merger ($\gtrsim$viscous timescale $\sim 0.1-1 {\rm s}$), it can deposit a large fraction of this rotational energy into the kilonova ejecta through magnetic-dipole spindown. Whether or not the merger remnant remains stable long-enough for this to take place depends sensitively on the total mass of the binary and $M_{\rm TOV}$. The mass of merging binary NSs is measurable for mergers detected by LIGO, however is inaccessible based on short GRB observations alone.
Thus, predictions for the energy of merger ejecta associated with short GRBs depend on the assumed mass-distribution of BNS mergers and the uncertain value of $M_{\rm TOV}$.

\begin{figure}[t]
\centering
  \includegraphics[width=0.47\textwidth,trim={0cm 0cm 0cm 0cm}]{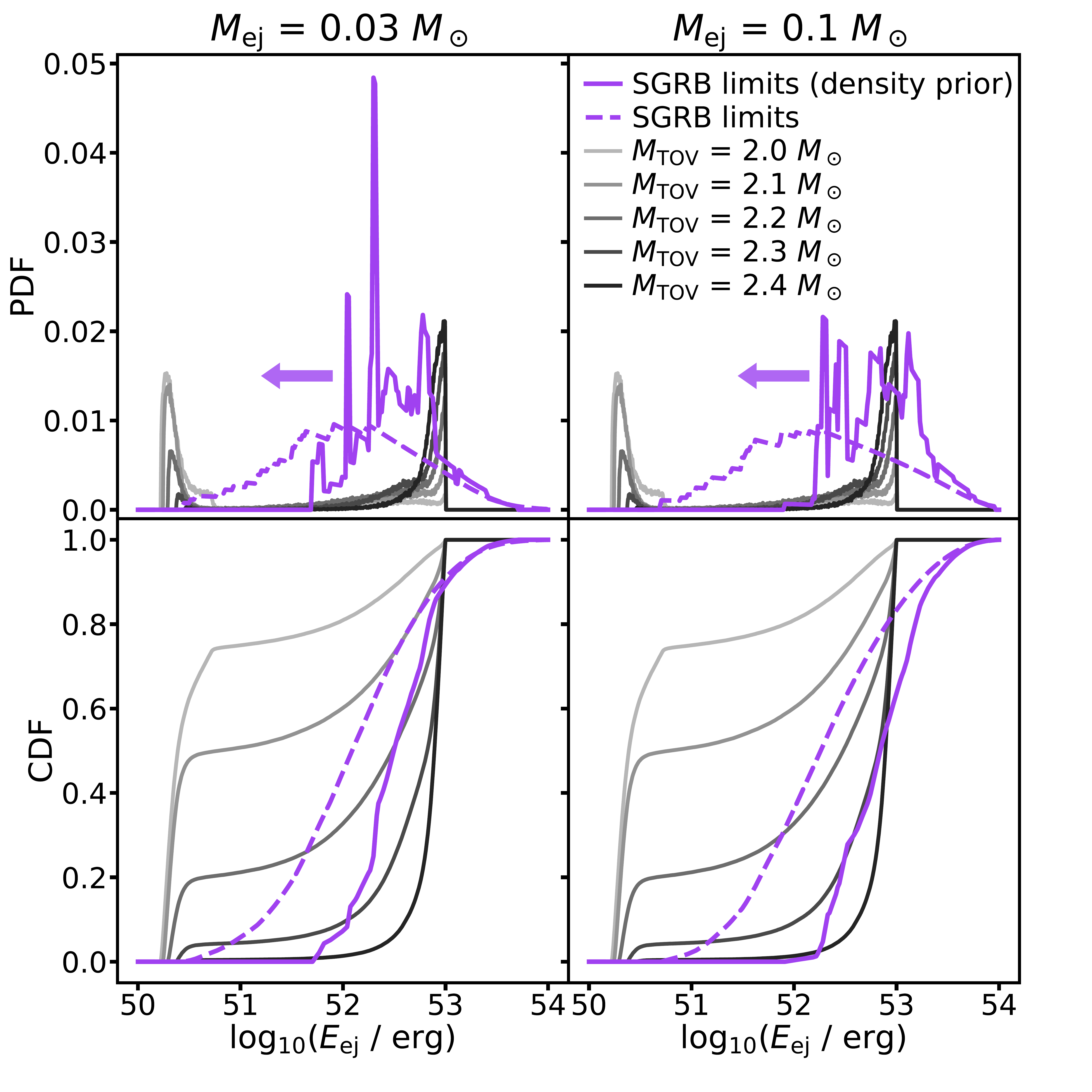}
  \vspace{-0.15in}
  \caption{{\it Top:} Probability distribution functions for the upper limits on the kinetic energies, $E_{\rm ej}$, allowed by radio observations of 27 short GRBs for ejecta masses of $M_{\rm ej}=0.03\,M_{\odot}$ (left) and $0.1\,M_{\odot}$ (right).
  PDFs represent an ``upper limit'' distribution and are shown for the sample with no density prior (dashed purple lines), in which case the shape of the distribution is mainly controlled by the choice of density grid and flux upper limits; and including the density prior (solid purple lines). Also shown are predictions for a range of $M_{\rm TOV}=2.0-2.4\,M_{\odot}$ (gray-scale lines). {\it Bottom:} Cumulative distribution functions for the observed and predicted distributions. }
  \label{fig:epdf}
\end{figure}

In Figure~\ref{fig:epdf} we compare the probability distribution of maximum energy $E_{\rm ej,max}$ obtained from our radio upper-limits (purple) to model-predicted probability distributions of $E_{\rm ej}$ (gray) created following \cite{MargalitMetzger2019}. We assume the cosmological population of BNS mergers follows the Galactic double NS distribution \citep{Kiziltan+13}. We then draw random NS masses from this distribution and estimate the ejecta mass and velocity for each pair $\{m_1,m_2\}$ based on fitting formulae to numerical-relativity simulations \citep{Coughlin+18}. In addition to the dynamical ejecta, we add a disk-wind ejecta component whose velocity is taken as $0.15c$ and whose mass is taken to be a fraction $0.4$ of the disk mass \citep{SiegelMetzger17}, estimated from \cite{Coughlin+19} (see also \citealt{Radice+18}). In the above we adopt a NS radius of $11 \, {\rm km}$ \citep{Capano+20}, but our results are not particularly sensitive to this choice.

The procedure above defines a PDF of kilonova ejecta energy lacking any additional magnetar energy deposition. Such energy deposition is then taken into account as a function of the merger remnant mass ($M_{\rm rem}^{\rm b} = m_1^{\rm b}+m_2^{\rm b}-M_{\rm ej}^{\rm b}$, where superscript ``b'' denotes baryonic mass) with respect to $M_{\rm TOV}$. We approximate this function as
$T = T_{\rm max} \left( M_{\rm rem}^{\rm b}/M_{\rm TOV}^{\rm b} \right)^\alpha$ for $M_{\rm rem}^{\rm b} \leq M_{\rm TOV}^{\rm b}$ and $T = T_{\rm max} \left[ \left( \xi - M_{\rm rem}^{\rm b}/M_{\rm TOV}^{\rm b} \right) / \left( \xi-1 \right) \right]^\beta$ for $M_{\rm rem}^{\rm b} > M_{\rm TOV}^{\rm b}$, with $T_{\rm max} = 10^{53} \, {\rm erg}$, $\xi = 1.18$, $\alpha = 2.35$, and $\beta = 1.3$. These parameters were found to reasonably approximate the extractable magnetar rotational-energy for a range of EOS tested with the {\tt RNS} code \citep{StergioulasFriedman95}, and ensure that $T=0$ (no magnetar energy deposition) for a NS that collapses before reaching solid-body rotation ($M_{\rm rm}^{\rm b} \gtrsim \xi M_{\rm TOV}^{\rm b}$).
For systems with long-lived magnetar remnants we add the extractable rotational-energy $T$ to the total kilonova ejecta energy, essentially assuming an efficiency $\zeta=1$ for this energy deposition process.
As illustrated in Figure~\ref{fig:epdf}, the resulting predicted distributions for all values of $M_{\rm TOV}$ is bimodal, with $E_{\rm ej}>10^{52.5}$~erg, representing ``long-lived'' remnants with magnetar energy deposition and $E_{\rm ej} \lesssim 10^{50.5}$~erg representing ``normal'' kilonova ejecta energies (without a magnetar energy boost). We note that these $E_{\rm ej}$ model predictions do not depend on radiation physics nor in particular on the microphysical parameters $\epsilon_e$, $\epsilon_B$, $p$, and $\zeta_e$.

For $M_{\rm ej}=0.03\,M_{\odot}$, the majority of the observational energy constraints are concentrated in the region $E_{\rm ej} \approx 10^{51.5}-10^{53.5}$, with a median upper limit of $\approx 10^{52.5}$~erg. A basic comparison of the existing short GRB limits to expectations demonstrates that we can rule out the majority of $M_{\rm TOV} = 2.3-2.4\,M_{\odot}$ models for $M_{\rm ej} = 0.03\,M_\odot$. Moreover, the limits rule out the maximum expected energy outputs for $M_{\rm TOV} = 2.2\,M_{\odot}$ and $2.3\,M_{\odot}$ models. The limits are less stringent in the case of $M_{\rm ej} = 0.1\,M_\odot$, however in this regime we can still rule out the majority of the $2.4\,M_\odot$ model.

\subsection{The Role of High-mass BNS Mergers}
A major assumption in the estimates of the previous subsection is that short GRBs are produced exclusively by BNS mergers which track the Galactic double NS mass distribution.
The recent detection of GW190425 by the LIGO and Virgo GW detectors \citep{gw190425} is a clear counter-example to this assumption, as its total mass is 5$\sigma$ above the mean Galactic NS binary population (though there is no observational proof that GW190425-like events produce short GRBs).
If a large fraction of mergers involve high-mass systems akin to GW190425 then the constraints on $M_{\rm TOV}$ would be weakened. This is due to the fact that only mergers with a total mass $\lesssim \xi M_{\rm TOV}$ can produce long-lived magnetar remnants capable of enhancing the ejecta energy. The effect would be similar if a large fraction of short GRBs originated from NS-BH mergers (e.g., \citealt{Gompertz2020}).

To investigate this in a more quantitative way, we add a population of ``high-mass'', GW190425-like BNS systems to our analysis. We take the mass of these systems to be $m_1=m_2=1.65 M_\odot$ motivated by GW190425 \citep{gw190425}, and parameterize the fractional contribution of this high-mass population to the total BNS population as $f_{\rm GW190425}$. 
This is similar to the approach recently adopted by \cite{Sarin+20}.
In this notation, $f_{\rm GW190425}=0$ corresponds to a purely Galactic double NS mass distribution, while $f_{\rm GW190425}=1$ corresponds to one with only high-mass GW190425-like systems (clearly at odds with GW170817).
We then perform a series of calculations for a grid of $M_{\rm TOV}$ and $f_{\rm GW190425}$ values: for each set of parameters we produce a model CDF for $E_{\rm ej}$ (as in Figure~\ref{fig:epdf}, but with the addition of the high-mass GW190425-like population); the conditional probability of the model parameters given the data ($E_{\rm ej}$ upper-limits summarized in Table~\ref{tab:param}) is then calculated as 
\begin{equation}
\label{eq:P_MTOV_f}
    P\left( M_{\rm TOV},f \vert {\rm data} \right) =
    \frac{P\left( {\rm data} \vert M_{\rm TOV},f \right)}{\int \int P\left( {\rm data} \vert M_{\rm TOV}^\prime,f^\prime \right) dM_{\rm TOV}^\prime df^\prime} ,
\end{equation}
where
\begin{equation}
\label{eq:P_data}
    P\left( {\rm data} \vert M_{\rm TOV},f \right) = \prod_{i} P\left( \left. E_{\rm ej} < E_{\rm ej,max}^{(i)} \right\vert M_{\rm TOV},f \right)  ,
\end{equation}
and we have introduced the shorthand notation $f \equiv f_{\rm GW190425}$. Equation~(\ref{eq:P_MTOV_f}) implicitly assumes a uniform prior $P(M_{\rm TOV},f) = constant$. Finally, the index $i$ runs over all sources in Table~\ref{tab:param} and the term $P( E_{\rm ej} < E_{\rm ej,max}^{(i)} \vert M_{\rm TOV},f )$ in Equation~(\ref{eq:P_data}) is calculated using the model CDFs.

\begin{figure*}
    \centering
    \includegraphics[width=0.7\textwidth]{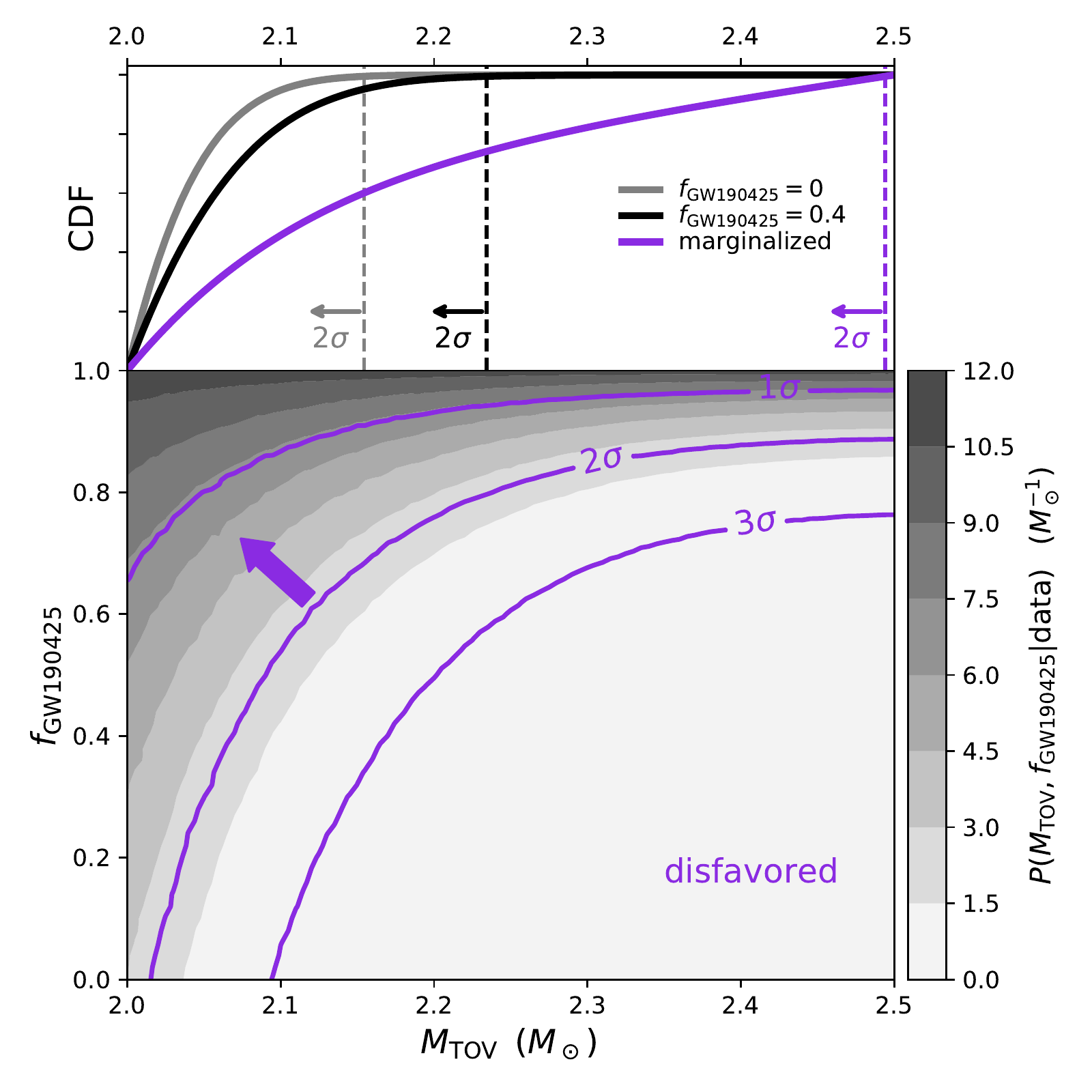}
    \vspace{-0.2in}
    \caption{{\it Bottom:} Joint PDF of the NS maximum mass, $M_{\rm TOV}$, and the fractional contribution $f_{\rm GW190425}$ of ``high-mass'' systems to the total population of BNS mergers (the remainder, $1-f_{\rm GW190425}$, of which is assumed to follow the Galactic BNS mass distribution). Mergers whose total mass is $\lesssim \xi M_{\rm TOV} \approx 1.2 M_{\rm TOV}$ are predicted to form ``long-lived'' magnetar remnants that would deposit a large amount of energy into the surrounding kilonova ejecta, a scenario that can be constrained by our radio observations (Table~\ref{tab:param}; Figure~\ref{fig:Evsnmej0.03}).
    Our radio upper limits can be reconciled with these predictions if few mergers produce long-lived magnetar remnants, implying small $M_{\rm TOV}$ and/or 
    moderate to large values for $f_{\rm GW190425}$ (purple contours, bottom panel). {\it Top:} Cumulative $M_{\rm TOV}$ distribution function for $f_{\rm GW190425}=0$ (grey), $f_{\rm GW190425}=0.4$ (black), and marginalized over $f_{\rm GW190425}$ (purple). $2\sigma$ (statistical) upper-limits on $M_{\rm TOV}$ are marked with vertical dashed curves, from which we find $M_{\rm TOV} \lesssim 2.15~M_\odot$ ($2.23~M_\odot$) for $f_{\rm GW190425}=0$ ($0.4$).}
    \label{fig:MTOV}
\end{figure*}

Figure~\ref{fig:MTOV} shows the joint constraints on the $M_{\rm TOV}$--$f_{\rm GW190425}$ parameter-space, calculated using the $M_{\rm ej}=0.1 M_\odot$ limits in Table~\ref{tab:param}.
The probability density function $P\left( M_{\rm TOV},f \vert {\rm data} \right)$ is plotted in the bottom panel, where purple contours demarcate the region to the left of which $1\sigma$, $2\sigma$, and $3\sigma$ of the cumulative probability is contained.
As discussed above, the constraints on $M_{\rm TOV}$ are reduced for a larger fraction of high-mass mergers (large $f_{\rm GW190425}$) because such mergers do not form long-lived magnetars that can enhance the kilonova ejecta energy to $\gtrsim 10^{52} \, {\rm erg}$, comparable to our radio limits.
The top panel of Figure~\ref{fig:MTOV} shows the marginalized $M_{\rm TOV}$ CDF (purple), in addition to the CDF at fixed $f_{\rm GW190425}=0$ (grey) and $f_{\rm GW190425}=0.4$ (black). $2\sigma$ upper-limits\footnote{This corresponds to the statistical significance, and does not include possible systematics.} on $M_{\rm TOV}$ based on these CDFs are plotted with corresponding vertical dashed curves. For $f_{\rm GW190425}=0$ the limits implied by the model necessitate $M_{\rm TOV} \lesssim 2.15 M_\odot$.
This is slightly deeper than other current limits on $M_{\rm TOV}$, which is constrained by massive-pulsar observations to $M_{\rm TOV} \gtrsim 2.0 M_\odot$ \citep{Demorest+10,Antoniadis+13,Cromartie+20} and by observations of GW170817 to $M_{\rm TOV} \lesssim 2.2-2.3 M_\odot$ \citep{MargalitMetzger17,Shibata+19}.
However, larger values of $f_{\rm GW190425}$ relax this $M_{\rm TOV}$ constraint. The ratio between the median total LIGO-inferred rate of BNS mergers and the median rate of BNS mergers akin to GW190425 ($\mathcal{R_{\rm GW190425}}$) suggests $f_{\rm GW190425} \sim 0.4$, albeit with large uncertainties given the small sample size \citep{gw190425}.
For this value of $f_{\rm GW190425}=0.4$, we find $M_{\rm TOV} \lesssim 2.23 M_\odot$.
Larger values of $f_{\rm GW190425}$ would further weaken our constraint. Indeed, marginalized over $f_{\rm GW190425}$, we find unconstraining limits on $M_{\rm TOV}$.

Compared to direct pulsar-mass measurements or even multi-messenger constraints from GW170817, the methodology we have developed here for constraining $M_{\rm TOV}$ is far more sensitive to various systematic uncertainties.
This method is statistical in nature and relies on a description of the, a-priori unknown, mass distribution of merging NSs (e.g. \citealt{Lawrence+15}; and see e.g. \citealt{Lasky+14} for an analogous methodology in the context of the magnetar-model for short GRB X-ray plateaus). Here we have assumed that this distribution is a mixture of the observed Galactic double NS distribution and a population of GW190425-like events governed by a single free parameter $f_{\rm GW190425}$, however the true mass distribution can in principle differ significantly from this assumption. Future GW detections will shed light on the local-Universe population of merging NSs and can be used to improve this methodology in the future.

The parameterization of the kilonova ejecta energy $E_{\rm ej}$ as a function of binary parameters, and in particular the universal form of magnetar energy deposition $T$ that we adopt, can also quantitatively influence our upper-limits on $M_{\rm TOV}$ (as can assumptions about the ejecta mass, though we have here adopted a conservative (large) value of $M_{\rm ej}=0.1M_\odot$).

Finally, the efficiency $\zeta$ of coupling between the magnetar spin-down energy and the energy deposited into the ejecta (here assumed to be $\zeta = 1$) can also impact our results, as can a more conservative (lower) value of $\epsilon_B$. 
If $\zeta < 1$ and the remainder ($1-\zeta$) of the spin-down energy is emitted as an ultra-relativistic pulsar wind, this would amount to depositing some fraction of the energy into even higher-velocity material leading to more luminous light-curves and stronger constraints on magnetar formation.
This is a consequence of the fact that the early ($t < t_{\rm dec}$) light-curve scales strongly with outflow velocity (at late times the light-curve depends only on the total outflow energy and is thus independent of $\zeta$).
Only in a scenario where $1-\zeta$ of the spin-down energy is instead radiated ``silently'' through GW losses would our derived constraints be weakened. However this scenario requires an unusually large NS ellipticity or extremely low external dipole field \citep{Ai+18}, which is not expected \citep[e.g.][]{Kiuchi+14}. Systematics in the numerical-relativity fitting formulae and approximate magnetar rotational-energy curve we have adopted can also impact our results, though the former do not have much affect on the high-$E_{\rm ej}$ distribution most relevant for this study.

\section{Observational Prospects in Constraining BNS Remnants}
\label{sec:obsprospects}

For our low-redshift sample, we find maximum ejecta energies of $E_{\rm ej, max}=(0.6-6.7)\times 10^{52}$~erg for an ejecta mass of $0.03 M_{\odot}$ and $E_{\rm ej, max}=(1.8-17.6)\times 10^{52}$~erg for $0.1 M_{\odot}$. This excludes GRB\,050509B for which there is only a density upper limit, and thus we calculate a wide range of $E_{\rm ej,max}\approx (0.7-39.7) \times 10^{52}$~erg for that burst. 
This work provides a re-analysis of seven short GRB observations presented in \citet{Fong2016}, and supercedes that analysis. The effects of more well-motivated assumptions on the value of $\epsilon_B$, combined with our new models which include relativistic effects, overall result in deeper constraints on $E_{\rm ej,max}$ by factors of $\approx 2-12$ for most of these events. For GRB\,090515, we provide a new analysis based on a fiducial density and for GRB\,130603B, we use both of the observations presented here and in \citet{Fong2016} and take the more constraining of the two solutions.
We use the 5.5~GHz VLA and ATCA limits from \citet{Klose2019}, who were searching for optically obscured star formation in their host galaxies, for an additional 10 GRBs and find $E_{\rm ej,max}=(2.1-13.1)\times10^{52}$~erg for $M_{\rm ej} = 0.03 M_\odot$ and $(2.5-27.4)\times10^{52}$~erg for $M_{\rm ej} = 0.1 M_\odot$. Our light curve model is nearly identical to that presented by \citet{Liu2020}. Our modeling overall provides more constraining limits on ejecta energies than studies which used the analytic approximation \citep{NakarPiran2011}, and the models primarily differ in their pre-peak behavior. The claim by \citet{Liu2020} that their model is significantly less constraining at late times than \citet{Fong2016} appears to be due to assumptions they make about the analytic model. In summary, the constraints on ejecta energies here are significantly deeper than any previous study on short GRBs.

Going forward, there are three primary observational avenues to place constraints on the nature of the remnants produced in BNS mergers: (i) continued observations of cosmological short GRBs, (ii) follow-up of GW events, and (iii) searches in untargeted radio surveys, such as the Very Large Array Sky Survey (VLASS; \citealt{Lacy2020}). We briefly explore prospects for each of these avenues here.

Our low-redshift sample has demonstrated that for $z\lesssim 0.5$, current deep VLA observations which reach depths of $F_{\nu}\approx 10-20\,\mu$Jy can achieve limits of $E_{\rm ej,max}\approx 10^{52}$~erg, ruling out indefinitely stable remnants, and potentially some supramassive remnants. As demonstrated in Figure~\ref{fig:epdf}, the predicted energy distribution is bimodal, with a low-energy peak at $E_{\rm ej}<10^{51}$~erg corresponding to kilonovae ejecta whose energy is not enhanced by magnetar energy deposition. In these cases, the peak flux density is $F_{\nu}\lesssim 0.1-1\,\mu$Jy at $\sim$GHz frequencies (for $z=0.1-0.4$), infeasible with current or planned radio facilities. Thus, while continued follow-up of short GRBs is an effective way to place limits of $E_{\rm ej} \approx 10^{52}$~erg, it is not possible to constrain energies well below these values from the cosmological short GRB population.

The remnants of short GRBs can in principle be constrained by observations at $\gamma$-ray and X-ray wavelengths. For instance, anomalous X-ray behavior that deviates from the standard GRB model, such as extended emission extending tens to hundreds of seconds in $\gamma$-ray light curves \citep{Bucciantini2012,Gompertz2013,Murase2018}, X-ray plateaus \citep{Rowlinson2013,Lu2015, Stratta2018}, and late-time X-ray excess emission on timescales of $\gtrsim 1$~day \citep{Perley2012,Fong2014} can all be interpreted as energy injection from long-lived magnetar remnants,
and under this assumption can be analogously used to constrain $M_{\rm TOV}$ (e.g. \citealt{Fan+13,Lasky+14,Gao+16,Sarin+20}).
The inferred magnetar energies from these injection sources are consistent with or slightly lower than those inferred from radio limits, with $\approx (0.3-5) \times 10^{52}$~erg. Among the current sample with radio observations, 13 events have anomalous X-ray behavior \citep{Rowlinson2013,Lu2015,Fong2016,Lien2016}, and 7 of these have constraints well below the extractable energy from a stable remnant of $E_{\rm ej,max}<10^{53}$~erg (Table~\ref{tab:param}). For these bursts, the inferred maximum energies from radio observations are comparable to the inferred injection energies from the X-ray band.

We can use the short GRB population to quantify the fraction of indefinitely stable magnetar remnants (e.g., those which can effectively transfer $10^{53}$~erg of energy to the environment), $f_{\rm stable}$. The sample presented here covers {\it Swift} short GRBs discovered over 2005-2016, out of a total detected 115 events (including 12 short GRBs with extended emission; \citealt{Lien2016}). For $M_{\rm ej} = 0.03\,M_{\odot}$, we derive $E_{\rm ej, max}<10^{53}$~erg (corresponding to the maximum energy deposited by an indefinitely stable magnetar with $M \lesssim M_{\rm TOV}$) for 22 short GRBs, or $19 \%$ of the total {\it Swift} short GRB population. Thus, we place a limit of $f_{\rm stable} < 0.81$.

\begin{figure}[t]
    \centering
    \includegraphics[width=0.45\textwidth]{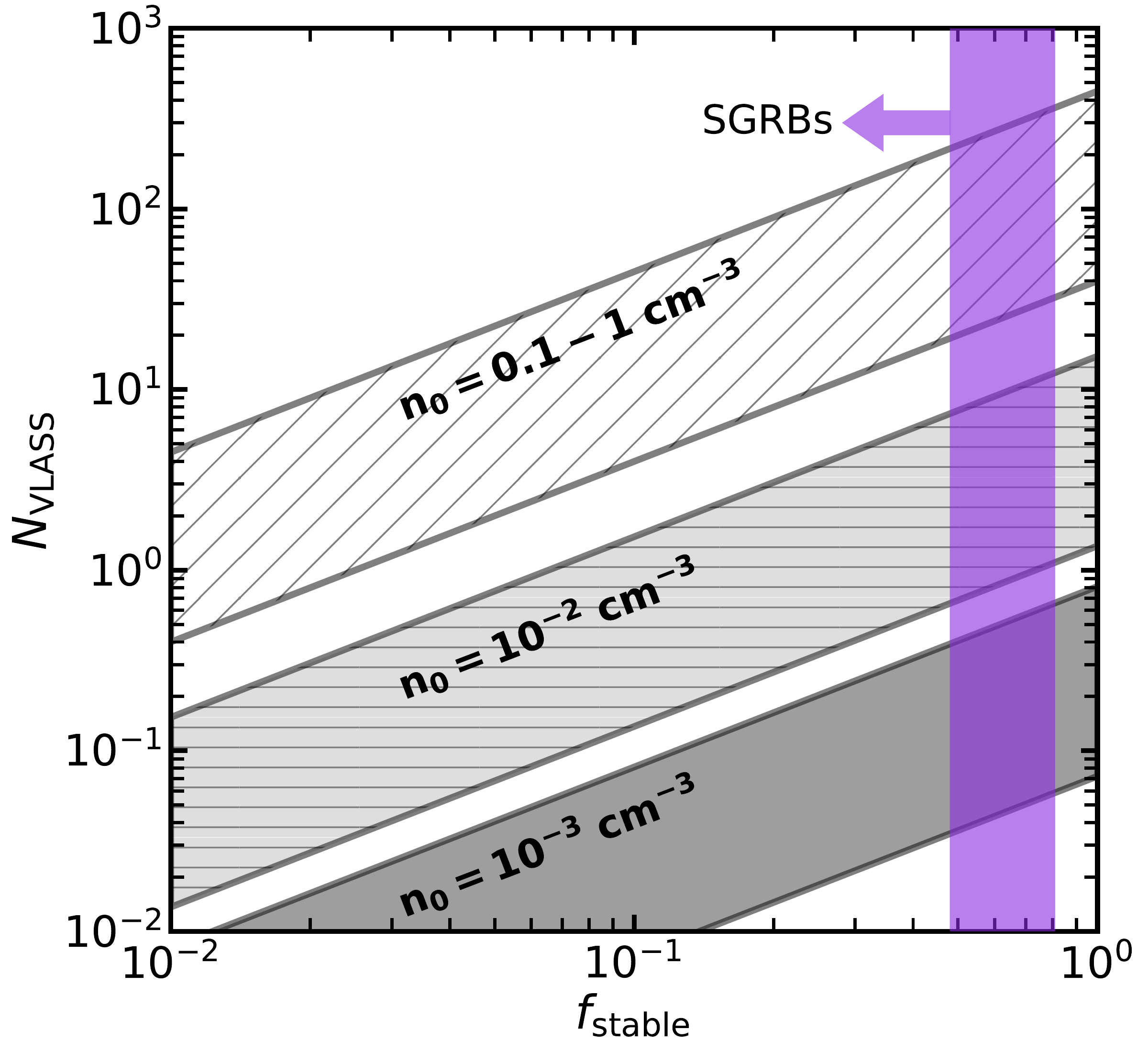}
    \caption{The number of radio magnetar remnants detectable at the level of $10\sigma$ in a VLASS epoch ($N_{\rm VLASS}$; $\sigma_{\rm RMS} = 120\,\mu$Jy) versus the fraction of stable remnants in the BNS merger population ($f_{\rm stable}$). Characteristic numbers are calculated for a stable remnant which deposits $E_{\rm ej}=10^{53}$~erg in the surrounding medium for varying densities ($n_0=10^{-3}-1$~cm$^{-3}$). The gray shaded and hatched regions for each density corresponds to the uncertainty in the BNS merger rate \citep{gw190425}. The purple shaded region corresponds to the range of $f_{\rm stable}$ determined by short GRBs. We note that $N_{\rm VLASS}$ corresponds to the detectability of radio remnants but does not include the contamination and uncertainties introduced by the proper identification of slowly-evolving radio transients as radio remnants.}
    \label{fig:nvlass}
\end{figure}

To date, every short GRB with radio follow-up observations on the relevant timescales has yielded non-detections, and the large majority have limits of $E_{\rm ej, max}<10^{53}$~erg (Table~\ref{tab:param}). In particular, if the low-redshift ($z<0.5$) subset of short GRBs is representative of the entire short GRB population in terms of the nature of the remnant, and there is no redshift dependence in remnant stability once a short GRB is produced, we can use this population to place a stronger constraint on the value of $f_{\rm stable}$. Of the 25 short GRBs at $z<0.5$, 19 have radio observations on the relevant timescales, while 13 events have $E_{\rm ej, max}<10^{53}$~erg, ruling out a stable remnant (Table~\ref{tab:param}). Thus, we place an additional constraint from the low-redshift short GRB population of $f_{\rm stable}<0.48$ (Figure~\ref{fig:nvlass}); this means that the majority of short GRBs do not produce stable remnants. This agrees with theoretical expectations (e.g., \citealt{MargalitMetzger2019}; see also \citealt{Piro+17}), although the exact requirements for post-merger product to form a disk and launch a GRB jet, as well as how this connects to the nature of the remnant is still an open question (e.g., \citealt{Margalit2015,Ruiz2017,Moesta+20}). 

This calculation assumes that BNS mergers are the progenitors of all short GRBs. If, for instance, NS-BH mergers contribute significantly to the short GRB population \citep{Gompertz2020}, the constraints on $f_{\rm stable}$ are weakened. Current volumetric rate estimates of NS-BH mergers are uncertain: a comparison of the upper limit on the rate of NS-BH mergers from GW facilities \citep{LIGO-O1O2} and the value measured from short GRBs \citep{Fong2015} demonstrates that current GW observations can accommodate any fraction of short GRBs produced by NS-BH mergers. Meanwhile, population synthesis studies result in rates that are comparable to (e.g., \citealt{Eldridge2019}) or considerably lower than \citep{Belczynski2020} the NS-NS merger population. The parameter space in which a typical $1.4\,M_{\odot}$ NS can be disrupted by a BH to produce an observable EM counterpart is extremely limited (\citealt{Foucart12,Capano+20}; e.g., low-mass and high-spin), and it is unlikely that NS-BH mergers are the majority progenitor population for short GRBs.

A second avenue to constrain the nature of the remnant is by more local GW events. Their proximity enables potentially deep constraints on the value of $E_{\rm ej, max}$ via EM observations or gravitational waves (e.g., \citealt{Murase2018}), although this is also dependent on the values of the microphysical parameters; in particular low values of $\epsilon_B$ weaken the constraints (e.g., Figure~\ref{fig:Evsnmej_epsB}; \citealt{Liu2020}). Indeed, GW170817 has a low inferred value of $\epsilon_B=0.0023$ \citep{Hajela2019}; from these, we infer $E_{\rm ej,max}=1.3 \times 10^{52}$~erg for the fast kilonova ejecta (Figure~\ref{fig:GW170817}). Several studies based on broad-band afterglow and in particular X-ray observations of GW170817, were able to rule out a long-lived stable or supramassive remnant \citep[e.g.][]{MargalitMetzger17,Shibata+17,Pooley+18,Margutti2018}. It has been argued that GW\,170817 produced a temporarily stable hypermassive NS \citep{MargalitMetzger17} and that this millisecond magnetar could explain the high ejecta mass, high velocity, and high electron fraction \citep{Metzger2018}. If the rate of low-mass BNS mergers (less massive than GW170817) is less then the rate of GW170817-like events, then a weak upper limit on $f_{\rm stable}$ also comes from the relative rates of GW170817-like events with respect to the total BNS merger rate (e.g., $f_{\rm stable}<0.6$) but with large uncertainties \citep{gw190425}; in reality, stable remnants are expected to only represent a small fraction of GW170817-like events. Observations of future GW events are another route to quantifying $f_{\rm stable}$ \citep{MargalitMetzger2019}, although the largest challenge will be the precise localization of these events via EM counterparts, and the increasing sensitivity of future GW facilities which will result in dimmer counterparts for normal events.

A final avenue to constrain the fraction of stable remnants is via detection in untargeted radio surveys. Short GRBs are highly collimated, and the observed rate depends sensitively on their opening angles \citep{Coward2012,Jin2018}. Thus, the short GRB rate is likely a factor of $\approx 100$ lower than the actual rate of BNS mergers \citep{Fong2015}. In contrast, the synchrotron radio signal from magnetar remnants is expected to be relatively isotropic, and thus every BNS merger which produces a stable remnant should in principle have an observable radio counterpart, regardless of the direction of the GRB jet. Therefore, radio remnants from stable magnetars should be bright radio signals and detectable in untargeted radio surveys, such as VLASS \citep{Metzger2015}, and in turn offer an additional constraint on $f_{\rm stable}$.  

Using the new light curve modeling presented here, we calculate the instantaneous rate of radio signals from magnetar remnants found in a single VLASS epoch \citep{Lacy2020}, $N_{\rm VLASS}$, by adapting Equation~8 from \citet{Metzger2015}. Given the fractional sky coverage of VLASS ($f_{\rm VLASS}=0.82$) and the typical RMS of a single VLASS epoch ($F_{\nu,{\rm VLASS}}=120\,\mu$Jy), this can be calculated as
\begin{equation}
N_{\rm VLASS} = f_{\rm VLASS} \times \frac{4\pi}{3}\left(\frac{L_\nu}{4 \pi F_{\nu,{\rm VLASS}}}\right)^{3/2} \mathcal{R} t_{\rm dur}.
\end{equation}
\noindent Here, $L_{\nu}$ is the spectral luminosity of the source in erg~s$^{-1}$~cm$^{-2}$~Hz$^{-1}$, $\mathcal{R}$ is the volumetric rate of BNS mergers in Gpc$^{-3}$~yr$^{-1}$, and $t_{\rm dur}$ is the survey timescale in years. To ensure a detection that is robust enough to detect variability, we fix $L_{\nu}$ to $1/10$ of the peak luminosity for a given model (in effect requiring a $10\sigma$ detection). For the value of $\mathcal{R}$, we employ the local volumetric rate of BNS mergers derived from LIGO\footnote{We note that this is not a 90\% confidence interval, as the calculation involves the union of rates assuming a uniform mass distribution, and that derived from GW170817 and GW190425, but does properly represent the uncertainty in the latest rate estimates.} ($\mathcal{R}$$ = 250-2810$~Gpc$^{-3}$~yr$^{-1}$; \citealt{gw190425}), and $t_{\rm dur}=3$~years \citep{Metzger2015}.

To determine $L_{\nu}$, we generate light curve models for $E_{\rm ej}=10^{53}$~erg (corresponding to stable remnants) at $\nu_{\rm VLASS}=3$~GHz for varying densities, $n_0 = 10^{-3}-1$~cm$^{-3}$. We show $N_{\rm VLASS}$ as a function of $f_{\rm stable}$ in Figure~\ref{fig:nvlass}, and note that the peak luminosity for the $0.1$ and 1~cm$^{-3}$ cases are similar. For low circumburst densities of $n_0 \approx 10^{-3}$~cm$^{-3}$, only $\approx 0.1-1$ are expected to be detected in VLASS in the most optimistic case where $f_{\rm stable}=1$. On the other hand, if BNS mergers overall originate in higher density environments (e.g. \citealt{Oconnor+20}; though see \citealt{Fong2015}), their radio remnants should be very luminous and on the order of $\approx 30-300$ should be detectable. Thus, comparing the VLASS-detected rate of radio remnants and the resulting constraint on $f_{\rm stable}$ with that from short GRBs will indirectly probe whether all BNS mergers produce short GRBs, and the connection between jet launching and remnant stability. In other words, if $\gtrsim 30-300$ radio remnants are identified in VLASS, this is in tension with constraints from short GRBs and will indicate that the short GRB population is biased. Meanwhile, the non-detection of any radio remnants in VLASS will constrain the phase space of environment density and $f_{\rm stable}$, more so than is possible with cosmological short GRBs. Finally, we note the important caveat that $N_{\rm VLASS}$ only represents detectable number in VLASS. The determination of whether or not a source is variable and can be properly identified as a remnant against the variable radio sky will necessarily be $<N_{\rm VLASS}$ \citep{Metzger2015}.

\section{Summary and Conclusions}
\label{sec:summaryconclusion}

We have presented new VLA radio limits of nine low-redshift short GRBs which provide the most constraining power from the cosmological short GRB population on energy deposited by magnetars in kilonova ejecta. We have also presented a uniform re-analysis of all available radio limits at $5.5-6.0$~GHz, alongside new light curve modeling, which is an improvement on past models by properly incorporating relativistic effects and the deep-Newtonian regime. We come to the following conclusions:

\begin{itemize}
    \item With the low-redshift sample, we can place limits on the maximum ejecta energies of $\approx\!0.6-6.7 \times 10^{52}$~erg, assuming an ejecta mass of $0.03 M_\odot$. 
   This increases to $\approx 1.8-17.6 \times 10^{52}$~erg when considering larger ejecta masses of $0.1\,M_{\odot}$. For GW170817, we obtain a limit on the energy in the fastest kilonova ejecta of $E_{\rm ej, max}=1.3 \times 10^{52}$~erg. Including all literature data, we find that the fraction of short GRBs that create stable magnetars must not exceed $\approx0.5-0.8$.
    \item We present predictions for the ejecta energies for varying $M_{\rm TOV}$, finding a bimodal distribution with peaks at $>10^{52.5}$~erg, corresponding to indefinitely stable or a fraction of supramassive magnetar remnants, and peaks at $<10^{50.5}$~erg, corresponding to the ejecta that are not boosted by magnetar energy deposition (i.e. cases where the remnant is a hypermassive NS or undergoes prompt collapse to a BH). Our short GRB limits consistently rule out indefinitely stable magnetars which deposit $10^{53}$~erg, and a fraction of supramassive NS cases.
    \item Assuming the BNS merger mass distribution follows the Galactic distribution, the radio limits constrain $M_{\rm TOV}<2.15\,M_{\odot}$,
    at $2\sigma$ confidence (statistical significance), 
    slightly deeper than existing constraints. Motivated by the discovery of the high-mass BNS merger GW190425, an increasing fraction of high-mass mergers quickly weakens these constraints. We find a less stringent constraint of $M_{\rm TOV}<2.23\,M_{\odot}$ 
    ($2\sigma$) 
    assuming a contribution of $40\%$ high-mass mergers to the current population, and unconstraining limits on $M_{\rm TOV}$ when marginalized over this uncertain fraction. Our predictions can be compared to the true fraction of high-mass mergers as it is solidified with future GW detections.
    \item We find that if current radio surveys such as VLASS were to detect tens to hundreds of stable magnetar remnants from BNS mergers, then this would imply that most short GRB remnants would be stable, which is at odds with current observations.  
\end{itemize}

Going forward, a concerted effort to uncover radio remnants in surveys (e.g., VLASS, \citealt{Lacy2020}; ASKAP/VAST \citealt{Murphy2013}, MeerKAT/ThunderKAT \citealt{Fender2016}), in parallel with dedicated follow-up observations of local BNS mergers (detected via GWs or low-redshift short GRBs) will help to constrain the fraction of stable remnants. In particular, the rate of detection of luminous radio remnants compared to the constraints from short GRBs will indirectly address whether short GRBs are a biased population, and how the launching of a successful GRB jet is connected to remnant stability. Moreover, as GW facilities delineate the true mass distribution of BNS mergers with additional discoveries, these can be used in conjunction with short GRB limits to place stringent constraints on the value of $M_{\rm TOV}$ and the NS equation of state.

\begin{acknowledgments}
G.S. acknowledges for this work was provided by the NSF through Student Observing Support award SOSP19B-001 from the NRAO. The Fong Group at Northwestern acknowledges support by the National Science Foundation under grant Nos. AST-1814782 and AST-1909358. B.M. is supported by NASA through the NASA Hubble Fellowship grant \#HST-HF2-51412.001-A awarded by the Space Telescope Science Institute, which is operated by the Association of Universities for Research in Astronomy, Inc., for NASA, under contract NAS5-26555. B.D.M. is supported in part by the Simons Foundation through the Simons Fellows Program (grant \# 606260).  K.D.A. is supported by NASA through the NASA Hubble Fellowship grant \#HST-HF2-51403.001-A awarded by the Space Telescope Science Institute, which is operated by the Association of Universities for Research in Astronomy, Inc., for NASA, under contract NAS5-26555.

The National Radio Astronomy Observatory is a facility of the National Science Foundation operated under cooperative agreement by Associated Universities, Inc. This work made use of data supplied by the UK Swift Science Data Centre at the University of Leicester. This research was supported in part through the computational resources and staff contributions provided for the Quest high performance computing facility at Northwestern University which is jointly supported by the Office of the Provost, the Office for Research, and Northwestern University Information Technology.  The LBT is an international collaboration among institutions in the United States, Italy and Germany. LBT Corporation partners are: The University of Arizona on behalf of the Arizona university system; Istituto Nazionale di Astrofisica, Italy; LBT Beteiligungsgesellschaft, Germany, representing the Max-Planck Society, the Astrophysical Institute Potsdam, and Heidelberg University; The Ohio State University, and The Research Corporation, on behalf of The University of Notre Dame, University of Minnesota and University of Virginia.
\end{acknowledgments}
\facilities{VLA, LBT:MODS}
\software{CASA \citep{CASA}, aoflagger \citep{Offringa2010,Offringa2012}, pwkit \citep{Pwkit}, RNS \citep{StergioulasFriedman95}}

\bibliographystyle{apj}
\bibliography{refs,journals_apj}

\end{document}